\documentclass[10pt]{article} 
\usepackage[utf8]{inputenc}
\usepackage{geometry}
\usepackage{authblk}
\usepackage{amsfonts}
\usepackage{xurl}
\usepackage{hyperref}
\usepackage{amsmath}
\usepackage{amsthm}
\usepackage{amsfonts}
\usepackage{dsfont}
\usepackage{pdflscape}
\usepackage{pgfplots}

\usepackage{mathrsfs}
\usepackage{breakurl}
\usepackage{indentfirst}
\usepackage{amssymb}
\usepackage{amssymb,amsmath,amsthm,mathrsfs,array,graphicx,bbm,enumerate,wasysym,dsfont}
\newtheorem{thm}{Theorem}[section]

\newtheorem{lem}[thm]{Lemma}

\usepackage{xcolor}
\theoremstyle{definition}
\newtheorem{defn}{Definition}[section]

\numberwithin{equation}{section}
\author{Xiao Han}
\affil{EPFL}
\date{}
\title{Infrared bound for the $O(N)$ model with Villain-type interaction on the torus of dimension $d\geq3$}
\allowdisplaybreaks
\begin{document}
\maketitle
\begin{abstract}
We prove the long-range order for the $O(N)$ model with Villain-type interaction on the torus with dimension $d\geq3$ in the low-temperature regime through the infrared bound. When $N=2$, this provides another proof for the long-range order of the Villain model. Our proof is suitable for more general positive definite rotation-invariant kernels on the sphere with some regularity conditions. The proof follows the classical strategy of Fröhlich, Simon and Spencer in 1976 \cite{FSS}, but replaces Fourier analysis on $\mathbb{R}$ with harmonic analysis on the sphere, where spherical harmonics play the role of the Fourier basis.
\end{abstract}
\section{Introduction}
The $O(N)$ model on $\mathbb{T}_L^d$, the torus with side length $L$ and dimension $d$, is a random configuration $\sigma:\mathbb{T}_L^d \to \mathbb{S}^{N-1}$ with probability measure:
\[
\mathbb{P}(d\sigma)\propto \prod_{z\sim w} K_\beta(\sigma_z,\sigma_w)d\sigma,
\]
where $d\sigma=\prod\limits_{z\in \mathbb{T}_L^d}d\sigma_z$ and $d\sigma_z$ is the normalized surface measure on the sphere $\mathbb{S}^{N-1}$, $\beta>0$ is the inverse temperature, and $K_\beta^{}: \mathbb{S}^{N-1} \times \mathbb{S}^{N-1} \to \mathbb{R}$ is symmetric and represents the interaction between the nearest neighbours. The most typical example is the spin $O(N)$ model \cite{Spin, Biskup, Bauer, Ueltschi}, where $$
K_\beta^{Spin}(x,y):=\exp\big(\beta \langle x,y\rangle\big).$$ 

The infrared bound for the spin $O(N)$ model on the torus of dimension $d\geq 3$ is a classical result given by Fröhlich, Simon and Spencer in 1976 \cite{FSS}, where they provided different approaches to the following result: 
\begin{equation}
\label{Infra}
 \sum_{z,w\in \mathbb{T}_L^d}\mathbb{E}_{Spin}\big[ \langle\sigma_z, \sigma_w\rangle  e^{2\pi i \cdot \frac{ (w-z)\cdot {\omega}}{L}}\big]\leq \frac{C_0(d)N|\mathbb{T}_L^d|}{\beta\Big[\sum_{l=1}^d[1-\cos (2\pi\cdot\frac{\omega_l}{L})]\Big]},
\end{equation}
for any $\omega=(\omega_1,\dots, \omega_d)\in \{1,2,\dots,L\}^d\setminus\{(L,\dots,L)\}$, where $C_0(d)>0$ is a constant only depends on $d$. By summing (\ref{Infra}) over all $\omega$, one will have long-range order of the spin $O(N)$ model for $d\geq 3$ in the low-temperature regime: there exist constants $C_{Spin}>0$ and $\beta_M^{Spin}(N,d, C_{Spin})>0$ such that for any $\beta>\beta_M^{Spin}$ and $L>0$ we have 
$$
\frac{1}{|\mathbb{T}_L^d|^2}\sum_{z,w\in \mathbb{T}_L^d}\mathbb{E}_{Spin}\langle \sigma_z,\sigma_w\rangle >C_{Spin}.
$$
And this is one of the key macroscopic behaviors of particular interest in these statistical mechanics models. In ferromagnetic spin systems, for instance, it implies the emergence of spontaneous magnetization. The infrared bound has also been applied in the Solid-On-Solid model to prove the localization, corresponding to the flat phase of a crystal surface \cite{BFL1982}. 

\medskip
The Villain-type interaction comes from the Villain model \cite{Villain1975}, which is a random configuration $\theta:\mathbb{T }_L^d \to [-\pi,\pi)$ with probability measure:
$$
\mathbb{P}_{Villain}(d\theta)\propto \prod_{z\sim w}\Big[\sum_{m\in\mathbb{Z}}\exp\Big(-\frac{\beta}{2}(\theta_z-\theta_w+2\pi m)^2\Big)\Big]d\theta,
$$ 
where  $\beta>0$ is the inverse temperature, $d\theta=\prod\limits_{z\in\mathbb{T}_L^d} d\theta_z$ and $d\theta_z$ is the Lebesgue measure on $[-\pi,\pi)$. Note that if we write $\sigma_z=(\cos \theta_z, \sin \theta_z)$, the XY model (Spin $O(2)$) will have interaction $K_\beta^{Spin}(\sigma_z,\sigma_w)=e^{\beta \langle\sigma_z, \sigma_w\rangle}=e^{\beta\cos (\theta_z-\theta_w)}$, which can be approximately seen as $ \sum\limits_{m\in \mathbb{Z}}\exp\big(-{\frac{\beta}{2}(\theta_z-\theta_w+2\pi m)^2}\big)$, as $\cos(\theta_z-\theta_w)= 1-\frac{1}{2}(\theta_z-\theta_w)^2+o(\theta_z-\theta_w)$ and if we ignore the constant $1$, which makes no influence on the model, the two interactions will have the same leading term. Since the $2\pi$-periodic property ($\theta_z \to \theta_z+ 2\pi$) is also kept under this transformation, the Villain model can be seen as an approximation of the XY model. Furthermore, from the Poisson summation formula, we have that
\begin{equation}
\label{1.2}
\sum\limits_{m\in \mathbb{Z}}\exp\big(-{\frac{\beta}{2}(\theta_z-\theta_w+2\pi m)^2}\big)=\frac{1}{\sqrt{2\pi \beta}}\sum_{k\in \mathbb{Z}} e^{-\frac{k^2}{2\beta}}\cdot e^{ik(\theta_z-\theta_w)},
\end{equation}
which is the discretization of the Gaussian law $\exp(-\frac{\beta}{2}\big(\theta_z-\theta_w)^2\big)=\frac{1}{\sqrt{2\pi \beta}}\int e^{-\frac{t^2}{2\beta}}e^{it(\theta_z-\theta_w)}dt$. Note that
$$
\partial_\tau \sum_{k\in\mathbb{Z}}e^{-k^2\tau}\cdot e^{ikx}=-\sum_{k\in\mathbb{Z}}k^2e^{-k^2\tau}\cdot e^{ikx}=\partial_x^2 \sum_{k\in\mathbb{Z}}e^{-k^2\tau}\cdot e^{ikx},
$$
the Villain action is exactly the heat kernel on $\mathbb{S}^1$ if we take $\tau=\frac{1}{2\beta}$ and $x=\theta_z-\theta_w$. Inspired by this, in the present paper we will consider the following model. 
\begin{defn}
\label{Big}
The $O(N)$ model with Villain-type interaction (Villain $O(N)$ model) on $\mathbb{T}_L^d$ is a random configuration $\sigma:\mathbb{T}_L^d \to \mathbb{S}^{N-1}$ with probability measure
$$
\mathbb{P}_{Villain}^N(d\sigma)\propto \prod_{z\sim w}h_\beta^{\mathbb{S}^{N-1}}(\sigma_z,\sigma_w) d\sigma,
$$
where $d\sigma$ is the product of normalized surface measure, $h_\beta^{\mathbb{S}^{N-1}}$ is the integral kernel of $e^{\frac{1}{2\beta}\Delta_{\mathbb{S}^{N-1}}}$ and $\Delta_{\mathbb{S}^{N-1}}$ is the Laplace-Beltrami operator on $\mathbb{S}^{N-1}$. 
\end{defn}
$h_\beta^{\mathbb{S}^{N-1}}$ is also the transition probability density of the Brownian Motion at time $\frac{1}{\beta}$ on the sphere. For $N=2$, this is exactly the Villain model as we can ignore the constant $\frac{1}{\sqrt{2\pi\beta}}$. This type of generalization has also been mentioned and discussed in \cite{Villain, DubedatFalconet}.

In this paper, we will prove an infrared-bound-type result to show the long-range order for this $O(N)$ model with Villain-type interaction in the low-temperature regime for $N\geq 2$. We first revisit and reformulate the arguments in the Appendix of \cite{FSS} in a way that makes the underlying structure more explicit. This reformulation allows us to extend the classical strategy from the setting of the Fourier transform on $\mathbb{R}$ to spherical harmonics and Fourier series on higher-dimensional spheres.
\section{Main results}
Before we introduce the result related to the discrete Fourier spectrum, let us look at some generalizations of Fourier analysis on $\mathbb{R}$. These two aspects not only follow the same proof philosophy, but also yield very similar results.

We will first introduce a certain family of potentials and prove the long-range order or localization for models with these potentials.
\begin{defn}
\label{conA}
We say that a potential $V:\mathbb{R} \to \mathbb{R}$ satisfies condition $A$ if and only if:
\begin{enumerate}
    \item There exists $p > 0$ with $\int p(t)\, e^{itx}\, dt = e^{-V(x)}$;
    \item $(\log p)'' \geq -M$ for some constant $M > 0$, i.e.,
    \[
    \frac{p''}{p} - \left(\frac{p'}{p}\right)^2 \geq -M;
    \]
    \item $p''\in L^1(\mathbb{R}).$
\end{enumerate}
\end{defn}
Note that condition 1 basically means that $e^{-V(x)}$ is positive definite; condition 2 means that $p$ does not decay faster than a Gaussian law. These conditions are satisfied for $V(x)=\beta|x|^\alpha$ with $0<\alpha\leq2$ and $\beta>0$. In particular, for $\alpha =2$, $p$ is the Gaussian law, and for $\alpha=1$, $p$ is the Cauchy law. In \cite{Ondual}, very close conditions are given, and there is a nice discussion on what kind of $V$ could satisfy them. We refer the interested reader to Example 1 and Appendix D of \cite{Ondual} for these potentials.

Now we consider a model on a $d$-dimensional torus  $\sigma: \mathbb{T}_L^d \to \mathbb{R}^N $($d \geq 3$) with Hamiltonian
\[
H(\sigma) = \sum_{z\sim w,\, j=1 \dots N} V(\sigma_z^j - \sigma_w^j),
\]
and probability measure
\[
\mathbb{P}_V(d\sigma)\propto \exp\big(-H(\sigma)\big)\prod_zd\nu(\sigma_z),
\]
where we denote $\sigma_z:=(\sigma_z^1, \dots,\sigma_z^N)$ and $\nu$ is any finite measure with finite second moment. The following theorem is a slight generalization of the long-range order of the spin $O(N)$ model.

\begin{thm}
\label{INF0}
If the potential $V$ satisfies condition $A$, then there exists a constant $C(N,d)>0$ that only depends on $N$ and $d$, such that
\begin{equation}
\label{2.1}
\frac{1}{|\mathbb{T}_L^d|^2}\sum\limits_{z, w}\mathbb{E}_V\|\sigma_z - \sigma_w\|^2 \leq C(N, d)M.
\end{equation}
\end{thm}
For the spin $O(N)$ model, we take $\nu$ to be the normalized surface measure on $\mathbb{S}^{N-1}$, $V(x)=\frac{\beta}{2}x^2$ and $M=\frac{1}{\beta}$, so that $$\mathbb{P}_V(d\sigma)\propto \exp\big(-\sum\limits_{z\sim w}\frac{\beta}{2}\|\sigma_z-\sigma_w\|^2\big)\propto \prod\limits_{z\sim w} \exp \big(\beta\langle \sigma_z,\sigma_w\rangle\big),$$
$p(t)=\frac{1}{\sqrt{2\pi \beta}}\exp(-\frac{t^2}{2\beta})$ and $(\log p)''=-\frac{1}{\beta}\geq-M$. We will have that
\begin{equation}
\label{Square}
2-2\cdot\frac{1}{|\mathbb{T}_L^d|^2}\sum\limits_{z, w}\mathbb{E}_V\langle\sigma_z, \sigma_w\rangle=\frac{1}{|\mathbb{T}_L^d|^2}\sum\limits_{z, w}\mathbb{E}_V\|\sigma_z - \sigma_w\|^2 \leq C(N, d)M=\frac{C(N,d)}{\beta}.
\end{equation}
This means that this is exactly the spin $O(N)$ model, and we can pick $0<C_{Spin}<1$ and $\beta_M^{Spin}(N,d,C_{Spin})>0$ so that for any $\beta>\beta_M^{Spin}$ and $L>0$ we have
\begin{equation}
\frac{1}{|\mathbb{T}_L^d|^2}\sum\limits_{z, w}\mathbb{E}_V\langle\sigma_z, \sigma_w\rangle >C_{Spin},
\end{equation}
which is the long-range order of the spin $O(N)$ model.

We also perturb this result a bit and consider another model $\sigma: \mathbb{T}_L^d \to \mathbb{D}$ ($d \geq 3$), where $\mathbb{D} = \mathbb{R} \text{ or } \mathbb{Z}$, with Hamiltonian
\[
H(\sigma) = \sum_{z\sim w } V(\sigma_z  - \sigma_w ),
\]
and probability measure
\[
\mathbb{P}_\mathbb{D}(d\sigma)\propto \exp\big(-H(\sigma)\big)\delta_{\sigma_{z_0}=0} d\sigma,
\]
where $z_0$ is a fixed vertex in the torus, $d\sigma=\prod\limits_z d\sigma_z$ and $d\sigma_z$ is Lebesgue measure if $\mathbb{D}=\mathbb{R}$; counting measure $\sum\limits_{n\in \mathbb{Z}}\delta_n$ if $\mathbb{D}=\mathbb{Z}$. 
\begin{thm}
\label{Ginz1}
If the potential $V$ satisfies condition $A$, then there exists a constant $C_1(d)>0$ that only depends on $d$, such that
\begin{equation}
\label{7799}
\frac{1}{|\mathbb{T}_L^d|^2}\sum\limits_{z, w}\mathbb{E_D}(\sigma_z - \sigma_w)^2 \leq C_1(d)M ,
\end{equation}
\end{thm}
The setting of this theorem fits very well with the Ginzburg-Landau $\nabla\phi$ interface model on the torus \cite{BLL1975, BL1976, Velenik}, which can be defined as a random configuration $\sigma: \mathbb{T}_L^d \to \mathbb{D}$, where $\mathbb{D} =\mathbb{R} \text{ or } \mathbb{Z}$, with probability measure:
$$
\mathbb{P}_{GL}(d\sigma) \propto \exp\big(-\beta\sum_{z \sim w}V_{GL}(\sigma_z-\sigma_w)\big)\delta_{\sigma_{z_0}=0} d\sigma.
$$
Thus, if we take $V=\beta V_{GL}$ that satisfies condition $A$, we will have (\ref{7799}), which implies the localization of the model \cite{BrydgesSpencer, MagazinovPeled, Dario2024, Sellke}, meaning that the height fluctuations of the random interface remain bounded even when $L\to \infty$. In particular, for $V_{GL}=|x|$, so that $V=\beta |x|$, this theorem will imply the localization of the Solid-On-Solid model, which is proved in \cite{BFL1982}.

For $\mathbb{D}=\mathbb{Z}$, a stronger and very similar result is shown in \cite{Ondual}. The two approaches followed slightly different Fourier duality but with the same philosophy. Our proofs for Theorem \ref{INF0} and Theorem \ref{Ginz1} follow the proof strategy of \cite{FSS} and \cite{BFL1982}.

Now we are ready to apply these techniques to the $O(N)$ model with Villain-type interactions, which is related to discrete Fourier spectrum.  We will first provide another proof of the long-range order of the Villain model on the torus of dimension $d\geq 3$ in the low-temperature regime. Recall that the Villain model on a $d$-dimensional torus ($d\geq 3$) is a random configuration $\theta:\mathbb{T }_L^d \to [-\pi,\pi)$ with probability measure:
$$
\mathbb{P}_{Villain}(d\theta)\propto \prod_{z\sim w}\Big[\sum_{m\in\mathbb{Z}}\exp\Big(-\frac{\beta}{2}(\theta_z-\theta_w+2\pi m)^2\Big)\Big]d\theta,
$$ 
where  $\beta>0$ is the inverse temperature, $d\theta=\prod d\theta_z$ and $d\theta_z$ is the Lebesgue measure on $[-\pi,\pi)$. For convenience, we write $\sigma_z=(\cos\theta_z, \sin \theta_z)$ and note that the Lebesgue measure on $[-\pi,\pi)$ still induces the uniform measure on $\mathbb{S}^1$.

\begin{thm}
\label{V0}
We have long-range order of the Villain model on the torus of dimension $d\geq 3$ in the low-temperature regime: there exist constants $C_{Villain}>0$ and $\beta_M(d,C_{Villain})>0$ such that for all $\beta>\beta_M$  and $L \geq 2$, 
\begin{equation}
\frac{1}{|\mathbb{T}_L^d|^2}\sum_{z, w}\mathbb{E}_{Villain}\langle\sigma_z,\sigma_w\rangle \geq C_{Villain}.
\end{equation}
\end{thm}
We prove it with the same strategy as Theorem \ref{INF0}, i.e., through the transfer matrix formalism and infrared bound. This result with more precise asymptotics of the spin correlations has been established using techniques based on polymer expansion \cite{FS1981, FS1982, DarioWu2024, DarioWu2025}. Our proof basically replaces the double commutator part (Lemma A.3) in the Appendix of \cite{FSS} by some analysis on the discrete Fourier basis on $\mathbb{S}^1$, which is just $\{e^{ikx}\}_{k\in\mathbb{Z}}$, and they appear naturally in the Villain kernel. This double commutator part will appear also in the proof of Theorem \ref{INF0}, and we will mention it in that part. When the Fourier basis is discrete, it turns out that the double commutator part will be much more complicated, as there will be some off-diagonal terms, and we have to replace some non-negative terms in the classical proof with non-negative definite matrices (or non-negative definite operators).

\medskip
Recall Definition \ref{Big} for the definition of the Villain $O(N)$ model. We will then prove the long-range order for this model.

\begin{thm}
\label{VN0}
We have long-range order for Villain $O(N)$ model on the torus of dimension $d\geq 3$ in the low-temperature regime: for $N\geq 3$, there exist constants $C^N_{Villain}>0$ and $\beta_M^N(d,C^N_{Villain})>0$, such that for all $\beta>\beta_{M}^N$ and $L \geq 2$,
\begin{equation}
\label{VN012}
\frac{1}{|\mathbb{T}_L^d|^2}\sum_{z, w}\mathbb{E}^N_{Villain}\langle\sigma_z,\sigma_w\rangle \geq C_{Villain}^N.
\end{equation}
\end{thm}
In principle, this proof is parallel to the one for the Villain model. We apply the Fourier analysis on the sphere, which is related to the spherical harmonics. In fact, there is a similar expansion of the heat kernel on the sphere as in (\ref{1.2})
\begin{equation}
h_\beta^{\mathbb{S}^{N-1}}(x,y)=\sum_{l=0}^\infty e^{-\frac{l(l+N-2)}{2\beta}}\sum_{m=1}^{d_l}Y_{l,m}(x)\overline{Y_{l,m}(y)},
\end{equation}
where $d_l=\frac{(2l+N-2)(l+N-3)!}{l! (N-2)!}$, $Y_{l,m}$ are the spherical harmonics and form an orthonormal basis of $L^2(\mathbb{S}^{N-1},d\mu,\mathbb{C})$ \cite{DaiXu}. Moreover, for any $l\geq0$, the family $\{Y_{l,m}\}$ forms an orthonormal basis of $\mathcal{H}_l$, the space of spherical harmonics of degree $l$ on the sphere. In the proof, we manage to find the non-negative matrices (non-negative operators) in this spherical-harmonic representation and use some properties of the spherical harmonics to recover the same key inequality as in the proof of the Villain model.

The proof also works if $d\sigma$ is replaced by any other finite positive measure on the sphere. Moreover, any continuous rotation-invariant kernel $K(x,y)=K_0\big(\langle x, y\rangle\big)\in L^2(\mathbb{S}^{N-1}\otimes\mathbb{S}^{N-1}, \mathbb{R})$ on $\mathbb{S}^{N-1}$ ($N\geq 3$) can be written in the form
\begin{equation}
\label{RI}
K(x,y)=\sum_{l=0}^\infty q_l\sum_{m=1}^{d_l}Y_{l,m}(x)\overline{Y_{l,m}(y)},
\end{equation}
where 
\begin{equation}
\begin{aligned}
q_l:= \frac{\Gamma\left(\frac{N}{2}\right)}
{\sqrt{\pi}\,\Gamma\left(\frac{N-1}{2}\right)}
\int_{-1}^{1}
K_0(t)
\frac{\mathcal{C}_l^{\frac{N-2}{2}}(t)}
{\mathcal{C}_l^{\frac{N-2}{2}}(1)}
(1-t^2)^{\frac{N-3}{2}} \, \mathrm{d}t,
\end{aligned}
\end{equation}
and $\mathcal{C}$ is the Gegenbauer polynomial \cite{Szego, NIST, DLMF}. Suppose $K>0$, we will also show that this proof strategy is suitable for any positive sequence $\{q_l\}$ satisfying
\[
\sup_{l\geq1}\Big|\log\frac{q_l^2}{q_{l-1}q_{l+1}}\Big|\to 0 \text{ and } \Big|\log\frac{q_1}{q_0}\Big|\to 0,
\] 
as $\beta \to \infty$ (see condition 2 in Definition \ref{conA}). In particular, it will provide another proof for the spin $O(N)$ model, in which case $K(x,y)=e^{\beta \langle x, y\rangle}$ and
\begin{equation}
q_l=
(\frac{2}{\beta})^{\frac{N}{2}-1} \Gamma(\frac{N}{2})
I_{l+\frac{N}{2}-1}(\beta),
\end{equation}
where $I_{\Tilde{\nu}}$ is the modified Bessel function of the first kind of order $\Tilde{\nu}$ \cite{NIST, DLMF}. We believe that these analyses can be useful in a broader setting of statistical mechanics models related to the sphere.

At the end of the paper, we present another framework to work on these theorems, which is used in \cite{Ondual}. In this framework, we can see that the two approaches to the infrared bound of the spin $O(N)$ model in \cite{FSS}, usually known as Gaussian domination and transfer matrix formalism (the one we follow in this paper), and the idea in \cite{Ondual} are actually very close. Our proof on Villain-type interaction directly comes from these understandings. Although for the convenience of writing, we did not choose this framework, we wish to share these understandings.

The rest of the paper is constructed in the following way: In Section \ref{Lemma section}, we first prove two lemmas, which correspond to the transfer matrix formalism in Appendix of \cite{FSS}, and we will use the lemmas in all these theorems. We will also point out what the double commutator part in Lemma A.3 of \cite{FSS} corresponds to in these lemmas. In Section \ref{SGA}, we will prove Theorem \ref{INF0} and Theorem \ref{Ginz1} using the lemmas. The proof of Theorem \ref{V0} about the Villain model is presented in Section \ref{APFT} and the proof of Theorem \ref{VN0} about the Villain $O(N)$ model is presented in Section \ref{LOF}. We also offer one possible way to define the basis of spherical harmonics $\{Y_{l,m}\}$ explicitly and show an equation that we would use in the proof of Theorem \ref{VN0} about this definition of $\{Y_{l,m}\}$ in Appendix \ref{APA}. The other framework of proving these theorems is shown in Appendix \ref{APB}.

\medskip
\textbf{Statement of AI use}
This project started in September 2025. All the proofs are done without direct AI use, except for the proof in Section \ref{APFT} on the Villain model ($N=2$). Indirect AI use includes literature review, proofreading, and checking classical theorems about Hilbert-Schmidt operators and spherical harmonics.

We wish to avoid direct AI use in the whole project, but while working on the Villain model, there was a beautiful accident. We want to share this experience, and we believe it will be a nice example of AI use in mathematical research. (The notation here is independent of other parts of the paper.)

We were trying to show that when $\beta\to \infty$,
$$\int\big|\sum\limits_{j=1}^L(e^{i\theta_1^j}-e^{i\theta_2^j})\big|^2\prod\limits_{j=1}^L h_\beta^{\mathbb{S}^1}(e^{i\theta_1^j},e^{i\theta_2^j})d\mu(\theta_1)d\mu(\theta_2)\leq O(\frac{L}{\beta})\int\prod\limits_{j=1}^L h_\beta^{\mathbb{S}^1}(e^{i\theta_1^j},e^{i\theta_2^j})d\mu(\theta_1)d\mu(\theta_2).$$
We want to prove it as in the proof of the spin $O(N)$ model, i.e. by comparing the operator in the matrix form directly in the PSD order. This is not complicated as the right side is a diagonal matrix. But by expanding, we almost show that this inequality cannot hold, and in this case, we almost believe that this method cannot work for the Villain model. To check it, we ask AI to generate a counterexample. What surprises us is that AI generates a counterexample for $L=2$, but we believe that by Minkowski's inequality there is a trivial bound in the form of 
$$\int\big|\sum\limits_{j=1}^L(e^{i\theta_1^j}-e^{i\theta_2^j})\big|^2\prod\limits_{j=1}^L h_\beta^{\mathbb{S}^1}(e^{i\theta_1^j},e^{i\theta_2^j})d\mu(\theta_1)d\mu(\theta_2)\leq O(\frac{L^2}{\beta})\int\prod\limits_{j=1}^L h_\beta^{\mathbb{S}^1}(e^{i\theta_1^j},e^{i\theta_2^j})d\mu(\theta_1)d\mu(\theta_2).$$

We explore the reason with AI, and we find out that in Minkowski's inequality we need to assume $\mu$ is a positive measure, which is not necessary for the proof by matrix comparison. And when we further ask AI if we can find a counterexample for $\mu$ positive, it directly gives us a proof. It's a technique that we will see in Section \ref{APFT}:
\begin{equation}
\label{AI}
\begin{aligned}
\int\big|\sum\limits_{j=1}^L(e^{i\theta_1^j}-e^{i\theta_2^j})\big|^2\prod\limits_{j=1}^L h_\beta^{\mathbb{S}^1}(e^{i\theta_1^j},e^{i\theta_2^j})d\mu(\theta_1)d\mu(\theta_2)&\leq O(\frac{1}{\beta})\int \sum_{j=1}^L e^{i(\theta_1^j-\theta_2^j)}\prod\limits_{j=1}^L h_\beta^{\mathbb{S}^1}(e^{i\theta_1^j},e^{i\theta_2^j})d\mu(\theta_1)d\mu(\theta_2)\\
&\leq O(\frac{L}{\beta})\int\prod\limits_{j=1}^L h_\beta^{\mathbb{S}^1}(e^{i\theta_1^j},e^{i\theta_2^j})d\mu(\theta_1)d\mu(\theta_2).
\end{aligned}
\end{equation}
In the first inequality, the right side is no longer the same matrix, but we can still compare the PSD order; in the second inequality, we need the fact that $\mu$ is a positive measure.

The matrix reshaping in Section \ref{APFT} is originally given by AI, and we rewrite it rigorously. The proof for $N\geq3$ is given by us, under the belief that there will be a somewhat similar structure to (\ref{AI}), which is provided originally by AI. We have used GPT-5.6 Sol in this process, while the whole manuscript is human-written.

\medskip
\textbf{Acknowledgements}
 {The author would like to thank Juhan Aru, Christophe Garban, Rémy Mahfouf, Nathan de Montgolfier, Sébastien Ott and Ron Peled for their valuable feedback and comments. This work is supported by the Eccellenza grant 194648 of the Swiss National Science Foundation. The author is also a member of NCCR SwissMAP.}
\section{Transfer Matrix Formalism, Infrared Bound and Long-range Order}
\label{Lemma section}
In this section, we want to construct a general framework, following the framework in Appendix of \cite{FSS}, that we can use for all the theorems in this article. So we will work on a general measure space $(X,\mathcal{F},\mu)$, where $X$ is a separable metric space, $\mathcal{F}$ is its Borel $\sigma$-algebra, and $\mu$ is a finite measure. In the rest of the paper, $X$ will be chosen to be $\mathbb{R}^N$ or $\mathbb{S}^{N-1}$. It could potentially be some Lie group like $SU(N)$ or more general Riemannian manifolds when we consider models on these spaces \cite{DallOlioZapata, Grigoryan}. We will also work on a general kernel $K:  X \times X  \to \mathbb{R^+}$ where $K$ is bounded, symmetric, and measurable. The setting of the spin $O(N)$ model and the Villain $O(N)$ model both satisfy these conditions. 

Let $L^2(X,\mu, \mathbb{C})$ be the separable Hilbert space equipped with inner product 
$$\langle f,g\rangle_\mu:=\int_Xf(x)\overline{g(x)}d\mu$$
for $f,g \in L^2(X,\mu, \mathbb{C})$. We have the following lemma:

\begin{lem}
\label{PDL1}
If we assume $K$ is non-negative definite: for any $g\in L^2(X,\mu, \mathbb{C})$,
\begin{equation}
\label{PD1}
\int_{X^2} g(x)\overline{g(y)}K(x,y)d\mu(x)d\mu(y)\geq 0,
\end{equation}
then for any $f\in L^2(X,\mu ,\mathbb{C})$, $L\geq 2$, $k=1,2,\dots L$, we have
\begin{equation}
\label{INE1}
\begin{aligned} 
& [1-\cos (2\pi\cdot\frac{k}{L})]\int_{X^L} \sum_{j=1}^{L}\Big[ f(x_1)\overline{f(x_j)} e^{2\pi i\cdot\frac{(j-1)k}{L}} \Big] \ K(x_1,x_2)\dots K(x_{L-1},x_{L})K(x_{L },x_{1})d\mu(x_1)\dots d\mu (x_{L})\\
\leq &  \int_{X^L} |f(x_1)-f(x_2)|^2 K(x_1,x_2)\dots K(x_{L-1},x_{L})K(x_{L },x_{1})d\mu(x_1)\dots d\mu (x_{L}).
\end{aligned}
\end{equation}
\end{lem}
\medskip
Before the proof of this lemma, we want to point out that (\ref{INE1}) is a naive version of the infrared bound on the $1d$ torus, in the sense that $x_1,\dots, x_L$ form a $1d$ torus of length $L$  and $K(\cdot ,\cdot)$ is the interaction between the nearest neighbours. Moreover, the right side of (\ref{INE1}) corresponds to the double commutator part in Lemma A.3 of \cite{FSS} on $1d$ torus. We will prove it by a standard trace analysis in the Hilbert space.

\medskip
\begin{proof}[Proof of Lemma \ref{PDL1}]
    Note that the operator $T:L^2(X,\mu,\mathbb{C})\to L^2(X,\mu,\mathbb{C})$ defined by
$$
Tg(y)=\int_X K(x,y)g(x)d\mu(x)
$$
is a Hilbert-Schmidt operator since $K$ is bounded and $\mu$ is finite so that $K \in L^2(X\times X,\mu \otimes\mu,\mathbb{R})$ \cite{ReedSimon, SimonTrace}. By the Hilbert-Schmidt theorem, we have a complete orthogonal basis $\{\phi_n\}_{n \geq 1}$ for $L^2(X,\mu,\mathbb{C})$ so that $T\phi_n=\lambda_n\phi_n$ with $\lambda_n\to 0$ as $n\to \infty$. Note that we have $\lambda_n \geq 0$ by taking $g=\phi_n$ in $(\ref{PD1})$, thus we can suppose that $\{\lambda_n\}$ are non-increasing and non-negative. 

Define the multiplication operator $F$ by $Fg(x):=f(x)g(x)$ and $\overline{F}$ by $\overline{F}g(x):=\overline{f(x)}g(x)$, and we write
$$r_{mn}:=\int f(x)\phi_m(x)\overline{\phi_n(x)}d\mu=\langle  \phi_m,\overline{F}\phi_n\rangle_\mu=\langle  F\phi_m,\phi_n\rangle_\mu,$$
so
$$\overline{r_{mn}} =\int \overline{f(x)} \overline{\phi_m(x)} {\phi_n(x)}d\mu=\langle  \phi_n, {F}\phi_m\rangle_\mu=\langle  \overline{F}\phi_n,\phi_m\rangle_\mu.$$
(Note that usually $r_{mn} \neq \overline{r_{nm}}$.)
By the fact that $f\in L^2(X,\mu, \mathbb{C})$, $\mu$ is finite and $K$ is bounded, we have
$$\int_{X^2}|f(y)|^2K(x,y)^2d\mu(x)d\mu(y)<\infty,$$
thus $FT$, $TF$, $T\overline{F}$ and $\overline{F}T$ are all Hilbert-Schmidt operators, so
\begin{equation}
\begin{aligned}
 Tr(TFT^{j-1}\overline{F}T^{L-j})=&\int_{X^L}   f(x_1)\overline{f(x_j)}  K(x_1,x_2)\dots K(x_{L-1},x_{L})K(x_{L },x_{1})d\mu(x_1)\dots d\mu (x_{L}) \\
 =& \sum_{n=1}^{\infty}\langle \phi_n ,TFT^{j-1}\overline{F}T^{L-j}\phi_n\rangle_\mu\\
 =&\sum_{m,n}\lambda_n^{L-j+1}\lambda_m^{j-1}r_{mn}\overline{r_{mn}}\\
 \geq  &  0,
\end{aligned}
\end{equation}
thus it is equal to 
\begin{equation} 
 \int_{X^L}  \overline {f(x_1)}{f(x_j)}  K(x_1,x_2)\dots K(x_{L-1},x_{L})K(x_{L },x_{1})d\mu(x_1)\dots d\mu (x_{L})  
 =\sum_{m,n}\lambda_m^{L-j+1}\lambda_n^{j-1}r_{mn}\overline{r_{mn}}. 
\end{equation}
And we need to prove that
\begin{equation}
\begin{aligned}
&\Big[1-\cos (2\pi\cdot\frac{k}{L})\Big]\sum_{j=1}^{L}\sum_{m,n}(\lambda_n^{L-j+1}\lambda_m^{j-1}+\lambda_{m}^{L-j+1}\lambda_n^{j-1})|r_{mn}|^2 e^{2\pi i\cdot \frac{(j-1)k}{L}}\\
\leq &2\sum_{m,n}(\lambda_n^L+\lambda_m^L-\lambda_n^{L-1}\lambda_m-\lambda_m^{L-1}\lambda_n)|r_{mn}|^2.
\end{aligned}  
\end{equation}
We can interchange the order of the summation since the traces are absolutely summable, so we only need to show that  

\begin{equation} 
 \Big[1-\cos (2\pi\cdot\frac{k}{L})\Big]\sum_{j=1}^{L} (\lambda_n^{L-j+1}\lambda_m^{j-1}+\lambda_{m}^{L-j+1}\lambda_n^{j-1} ) e^{2\pi i\cdot \frac{(j-1)k}{L}} 
\leq  2(\lambda_n^L+\lambda_m^L-\lambda_n^{L-1}\lambda_m-\lambda_m^{L-1}\lambda_n) . 
\end{equation}
This is the inequality in Lemma A.2 of \cite{FSS}.
\end{proof}
\medskip

Now, we work on the following model on a higher-dimensional torus. 
\begin{defn}
A general lattice model on $(X,\mathcal{F},\mu)$ with interaction kernel $K$ and periodic boundary condition is a random configuration 
$\sigma:{\mathbb{T}_L^d}\to X$ distributed with probability measure
\begin{equation}
   \mathbb{P}_{Lem}(d\sigma)\propto \prod_{z\sim w} K(\sigma_z,\sigma_w)\cdot \prod_z d\mu(\sigma_z).
\end{equation}
\end{defn}
We will show another lemma, where we directly use the result of Lemma \ref{PDL1} to prove an infrared-bound-type result and use it to find an upper bound for the fluctuations of the model. This lemma will be directly used when we prove most of our main theorems: Theorem \ref{INF0}, Theorem \ref{V0} and Theorem \ref{VN0}.

\begin{lem}
\label{PDL}
If we assume $K\otimes \dots \otimes K$ is non-negative definite: for any $g\in L^2(X^{{L}^{d-1}},\mu \otimes\dots\otimes\mu , \mathbb{C})$,
\begin{equation}
\label{PD2}
\int_{X^{{L}^{d-1}}}\int_{X^{{L}^{d-1}}} g(x )\overline{g(y )}\prod_{j}[K(x_j,y_j)d\mu(x_j)d\mu(y_j)]\geq 0,
\end{equation}  
then for any $L\geq 2$, $d\geq 3$, there exists a constant $c(d)>0$ such that for  $f\in L^2(X,\mu,\mathbb{C})$,
\begin{equation}
\label{3.9}
    \frac{1}{|\mathbb{T}_L^d|^2}\sum_{z,w}\mathbb{E}_{Lem} \big|f(\sigma_z)-f(\sigma_w)\big|^2\leq c \cdot M_f,  
\end{equation}
where $M_f:=\frac{1}{L^{d-1}} \max\limits_{\substack{s(\cdot)\in   \mathbb{S}^1 }}\mathbb{E}\Bigg[\Big|\sum\limits_{z'\in \mathbb{T}_L^{d-1}}s(z')\big[f(\sigma_{(1,z')})-f(\sigma_{(2,z')})\big]\Big|^2\Bigg]$ with $\mathbb{S}^1=\{e^{i\theta }|\theta \in [0,2\pi)\} $.
\end{lem} 
Before we prove it, we want to mention that if $X\in \mathbb{R}^N$, we can take $f(\sigma_z):=\sigma_z^j$ with $j=1,\dots,N$ in (\ref{3.9}). By summing over all the $j$, we will have a bound of  $\frac{1}{|\mathbb{T}_L^d|^2}\sum\limits_{z,w}\mathbb{E}_{Lem} \|\sigma_z-\sigma_w\|^2$, which is what we want in most cases. (See e.g. (\ref{Square}).) Note that every high-dimensional torus has a $1d$-torus structure if we see each hyperplane of the torus as one vertex, so Lemma \ref{PDL1} can be directly used and this is why $X^{L^{d-1}}$ appears. Moreover, $M_f$ corresponds to the double commutator part in Lemma A.3 of \cite{FSS}.
\medskip
\begin{proof}[Proof of Lemma \ref{PDL}]
Consider the space $X^{{L}^{d-1}}$, which is the configuration on the hyperplane of the torus. We rewrite $\sigma: \mathbb{T}^d_L \to X$ as $(\sigma_1,\sigma_2,\dots,\sigma_L)$, where $\sigma_j(\cdot):=\sigma_{(j,\cdot)}\in X^{{L}^{d-1}}$. Let $d\sigma_j= \prod\limits_{z'\sim w'}K\big(\sigma_j(z'),\sigma_j(w')\big)\prod\limits_{z'} d\mu\big(\sigma_j(z')\big)$, which is a finite measure. We use Lemma \ref{PDL1} in the separable Hilbert space $L^2(X^{{L}^{d-1}}, d\sigma_j ,\mathbb{C})$ with bounded kernel $\Tilde{K}(\sigma_{j_1},\sigma_{j_2}):=\prod\limits_{z'}K(\sigma_{j_1}(z'),\sigma_{j_2}(z')) $ on the $L^2$ function $\sum\limits_{z'}f(\sigma_j(z'))e^{-2\pi i\cdot\frac{z'\cdot \Tilde{\omega}}{L}}$, where $\Tilde{\omega}=(\Tilde{\omega_1},\dots,\Tilde{\omega_{d-1}})\in \{1,2,\dots,L\}^{d-1}$ (we can check (\ref{PD1})
by multiplying the finite factor $\prod\limits_{z'\sim w'}K\big(\sigma_{j_1}(z'),\sigma_{j_2}(w')\big)$ into $g$ in (\ref{PD2})) to get
\begin{equation}
\begin{aligned} 
&[1-\cos (2\pi\cdot\frac{k}{L})]\int\sum_{ j,z',w'} [f(\sigma_1(z'))\overline{f(\sigma_j(w'))}  e^{2\pi i \cdot \frac{(j-1)k+(w'-z')\cdot \Tilde{\omega}}{L}}]\mathbb{P}_{Lem}(d\sigma)\\ 
\leq&
\int|\sum\limits_{z'}f(\sigma_1(z'))e^{-2\pi i\cdot\frac{z'\cdot \Tilde{\omega}}{L}}-\sum\limits_{z'}f(\sigma_2(z'))e^{-2\pi i\cdot\frac{z'\cdot \Tilde{\omega}}{L}}|^2\mathbb{P}_{Lem}(d\sigma)\\ 
\leq&L^{d-1}M_f.
\end{aligned} 
\end{equation}
And we would get that for any $ \omega=(\omega_1,\omega_2,\dots,\omega_d)\in \{1,2,\dots,L\}^d$,
\begin{equation}
\label{lalala}
\begin{aligned} 
&[1-\cos (2\pi\cdot\frac{\omega_1}{L})]\int\sum_{  z,w} f(\sigma_z)\overline{f(\sigma_w)}  e^{2\pi i \cdot \frac{ (w-z)\cdot {\omega}}{L}}\mathbb{P}_{Lem}(d\sigma)\leq L^d M_f,
\end{aligned} 
\end{equation}
so
\begin{equation}
\begin{aligned} 
&\Big[\sum_{l=1}^d[1-\cos (2\pi\cdot\frac{\omega_l}{L})]\Big]\int\sum_{z,w} f(\sigma_z)\overline{f(\sigma_w)}  e^{2\pi i \cdot \frac{ (w-z)\cdot {\omega}}{L}}\mathbb{P}_{Lem}(d\sigma)\leq dL^{d}M_f,
\end{aligned} 
\end{equation}
and we have the infrared bound for $\omega\neq (L,\dots,L)$:
\begin{equation}
\begin{aligned} 
&\int\sum_{z,w} f(\sigma_z)\overline{f(\sigma_w)}  e^{2\pi i \cdot \frac{ (w-z)\cdot {\omega}}{L}}\mathbb{P}_{Lem}(d\sigma)\leq \frac{dL^{d}M_f}{\Big[\sum_{l=1}^d[1-\cos (2\pi\cdot\frac{\omega_l}{L})]\Big]}.
\end{aligned} 
\end{equation}
Summing over all $\omega\in \{1,2,\dots,L\}^d\setminus\{(L,\dots,L)\}$, note that  $\sum\limits_\omega e^{2\pi i \cdot \frac{ (w-z)\cdot {\omega}}{L}}=-1+L^d\cdot \mathds{1}_{w=z}$, we get
\begin{equation}
\begin{aligned} 
&\int\sum_{z,w} \frac{1}{2}|f(z)-f(w)|^2   \mathbb{P}_{Lem}(d\sigma)\leq \Big(\frac{1}{L^d}\sum_{\omega }\frac{1}{ \sum_{l=1}^d[1-\cos (2\pi\cdot\frac{\omega_l}{L})]}\Big) dL^{2d}M_f.
\end{aligned} 
\end{equation}
This implies that
\begin{equation}
\frac{1}{|\mathbb{T}^d_L|^2}\sum_{z,w}\mathbb{E}_{Lem}|f(\sigma_z)-f(\sigma_w)|^2\leq 2d\Big(\frac{1}{L^d}\sum_{\omega }\frac{1}{ \sum_{l=1}^d[1-\cos (2\pi\cdot\frac{\omega_l}{L})]}\Big) M_f.
\end{equation}
Note that there exists $c(d)>0$ such that $\frac{2d}{L^d}\sum\limits_{\omega }\frac{1}{ \sum_{l=1}^d[1-\cos (2\pi\cdot\frac{\omega_l}{L})]} < c$, since $\int_{[0,2\pi]^d}\frac{1}{ \sum_{l=1}^d[1-\cos  \theta_l ]}d\theta$ exists for $d\geq 3$ \cite{FSS}.
\end{proof}
\section{Models with Fourier Analysis on $\mathbb{R}$}
\label{SGA}
Now we switch back to the setting of Theorem \ref{INF0}, $\sigma: \mathbb{T}_L^d \to \mathbb{R}^N $($d \geq 3$) is a random configuration with Hamiltonian
\[
H(\sigma) = \sum_{z\sim w,\, j=1 \dots N} V(\sigma_z^j - \sigma_w^j),
\]
and probability measure
\[
\mathbb{P}_V(d\sigma)\propto \exp\big(-H(\sigma)\big)\prod_zd\nu(\sigma_z),
\]
where $\nu$ is any finite measure with finite second moment. For the convenience of the reader, we rewrite the statement of Theorem \ref{INF0}, and we will prove it by Lemma \ref{PDL}. Note that the proof idea is just to take $f(\sigma_z)=\sigma_z^j$ in Lemma \ref{PDL} and find a bound for $M_f$. This is the same strategy as in Appendix of \cite{FSS} and in \cite{BFL1982}.
\begin{thm}
\label{INF}
If the potential $V$ satisfies condition $A$, then there exists a constant $C(N,d)>0$ that only depends on $N$ and $d$, such that
\begin{equation}
\frac{1}{|\mathbb{T}_L^d|^2}\sum\limits_{z, w}\mathbb{E}_V\|\sigma_z - \sigma_w\|^2 \leq C(N, d)M.
\end{equation}
\end{thm}
\begin{proof}
Take $X=\mathbb{R}^N$, $d\mu=d\nu$, $K(x,y)=e^{-\sum\limits_{j=1}^NV(x^j-y^j)}$ for $x,y \in \mathbb{R}^N$ in Lemma \ref{PDL}; thus $X$ is a separable metric space, $\mu$ is finite, and $K$ is bounded, symmetric, and measurable. Note that $e^{-V}$ and $p$ are duals of each other, so they are both even and absolutely bounded by the value at $0$. Furthermore, $\int |p(t)|dt=\int p(t)dt=e^{-V(0)}<\infty.$ To verify (\ref{PD2}), we interchange the order of summation using Fubini’s theorem. Here $x=(x_1,\dots,x_m,\dots,x_{L^{d-1}})$ and $x_m\in\mathbb{R}^N$.
\begin{equation}
\label{PDH}
\begin{aligned}
&\int_{X^{{L}^{d-1}}}\int_{X^{{L}^{d-1}}} g(x )\overline{g(y )}\prod_m[K(x_m,y_m)d\mu(x_m)d\mu(y_m)]\\
=&\int_{X^{{L}^{d-1}}}\int_{X^{{L}^{d-1}}}g(x)\overline{g(y)}\prod_m \Bigg[\prod_j\Big[\int_{t_m^j}e^{it_m^j(x_m^j-y_m^j)}p(t_m^j)dt_m^j\Big]\cdot d\mu(x_m)d\mu(y_m)\Bigg]\\
=&\int_t \int_{X^{{L}^{d-1}}}\int_{X^{{L}^{d-1}}}g(x)\overline{g(y)}\prod_m \Bigg[\prod_j\Big[\int_{t_m^j}e^{it_m^j(x_m^j-y_m^j)}p(t_m^j)\Big]\cdot d\mu(x_m)d\mu(y_m)\Bigg]dt \\
=&\int_t\Bigg(\prod_{j,m} p(t_m^j)\Bigg)\cdot\Bigg(\int_{X^{{L}^{d-1}}}g(x)\prod_m\Bigg[\prod_j\Big[\int_{t_m^j}e^{it_m^j x_m^j}\Big]d\mu(x_m)\Bigg]\Bigg)\cdot \Bigg(\int_{X^{{L}^{d-1}}}\overline{g(y)}\prod_m\Bigg[\prod_j\Big[\int_{t_m^j}e^{-it_m^j y_m^j}\Big]d\mu(y_m)\Bigg]\Bigg)dt\\
=&\int_t \Bigg(\prod_{j,m} p(t_m^j)\Bigg)\cdot \Bigg|\int_{X^{{L}^{d-1}}}g(x)\prod_m\Bigg[\prod_j\Big[e^{it_m^j x_m^j}\Big]d\mu(x_m)\Bigg]\Bigg|^2 dt\\
\geq &0,
\end{aligned}
\end{equation}
where we write $t$ for all the $t_m^j$ and $dt=\prod dt_m^j$. Absolute integrability comes from the integrability of $|p|$ and the fact that $\mu$ is a finite measure and $g$ is an $L^2$ function:
\begin{equation}
\begin{aligned}
 &\int_t \int_{X^{{L}^{d-1}}}\int_{X^{{L}^{d-1}}}\Big|g(x)\overline{g(y)}\Big|\prod_m \Big[\prod_j[e^{it_m^j(x_m^j-y_m^j)}p(t_m^j)]\cdot d\mu(x_m)d\mu(y_m)\Big]dt\\
 =&\int_t \int_{X^{{L}^{d-1}}}\int_{X^{{L}^{d-1}}}\Big|g(x)\overline{g(y)}\Big|\prod_m \Bigg[\prod_j\Big| p(t^j_m)\Big|\cdot d\mu(x_m)d\mu(y_m)\Bigg]dt\\
 \leq &e^{-V(0)L^{d-1}N} \int_{X^{{L}^{d-1}}}\int_{X^{{L}^{d-1}}}\Big|g(x)\overline{g(y)}\Big|\prod_m \Big[ d\mu(x_m)d\mu(y_m)\Big]\\
 <&+\infty.
\end{aligned}
\end{equation}

One can see this as a reflection positivity: the product of a positive definite kernel will still be a positive definite kernel.
Now we take $f(\sigma_z)=\sigma_z^j$, note that $f$ is an $L^2$ function since $\nu$ has finite second moment. We aim to show that $M_f \leq M$, thus
\begin{equation}
\frac{1}{|\mathbb{T}^d_L|^2}\sum_{z,w}\mathbb{E}_V\|\sigma_z-\sigma_w\|^2=\frac{1}{|\mathbb{T}^d_L|^2}\sum_{j=1}^N\sum_{z,w}\mathbb{E}_V|\sigma_z^j-\sigma_w^j|^2\leq N\cdot c \cdot M.
\end{equation}
And we only need to take $C(N,d)=N\cdot c(d)$. 

Recall that $M_f:=\frac{1}{L^{d-1}} \max\limits_{\substack{s(\cdot)\in\mathbb{S}^1 }}\mathbb{E}\Bigg[\Big|\sum\limits_{z'\in \mathbb{T}_L^{d-1}}s(z')\big[f(\sigma_{(1,z')})-f(\sigma_{(2,z')})\big]\Big|^2\Bigg]$. Integration by parts shows that
\[
-ix \int p(t)\, e^{itx}\, dt = \int p'(t)\, e^{itx}\, dt,
\]
so for fixed $\sigma$, replace $x$ by $\sigma_{(1,z')}^j - \sigma_{(2,z')}^j$  and we will have

\begin{align}
\label{int1}
&\left|\sum_{\substack{z'}}s(z') \left(\sigma_{(1,z')}^j - \sigma_{(2,z')}^j\right)\right|^2
\prod_{z'}e^{-V\big(\sigma_{(1,z')}^j-\sigma_{(2,z')}^j\big)}\notag\\
=&\Bigg(\sum_{\substack{z'}}s(z') \left(\sigma_{(1,z')}^j - \sigma_{(2,z')}^j\right)\Bigg)\Bigg(\sum_{\substack{z'}}\overline{s(z')} \left(\sigma_{(1,z')}^j - \sigma_{(2,z')}^j\right)\Bigg)\prod_{z'}\Bigg[\int p(t_{z'}^j)e^{it_{z'}^j(\sigma_{(1,z')}^j - \sigma^{j}_{(2,z')})}dt^j_{z'}\Bigg]\notag\\
=&\Bigg(\sum_{\substack{z'}}s(z') \left(\sigma_{(1,z')}^j - \sigma_{(2,z')}^j\right)\Bigg)\Bigg(\sum_{\substack{z'}}\overline{s(z')} \left(\sigma_{(1,z')}^j - \sigma_{(2,z')}^j\right)\Bigg)\Bigg[\int \prod_{z'}p(t_{z'}^j)\cdot e^{it^j\cdot (\sigma_{(1,\cdot )}^j - \sigma^{j}_{(2,\cdot)})}dt^j \Bigg]\notag\\
=&i\Bigg(\sum_{\substack{z'}}\overline{s(z')} \left(\sigma_{(1,z')}^j - \sigma_{(2,z')}^j\right)\Bigg) \Bigg[\int \Big(\sum_{\substack{z'}}s(z') \partial_{t_{z'}^j}\Big)\prod_{z'}p(t_{z'}^j)\cdot e^{it^j\cdot (\sigma_{(1,\cdot )}^j - \sigma^{j}_{(2,\cdot)})}dt^j \Bigg]\\
=&-\int \Big(\sum_{\substack{z'}}\overline{s(z')} \partial_{t_{z'}^j}\Big)\Big(\sum_{\substack{z'}}s(z') \partial_{t_{z'}^j}\Big)\prod_{z'}p(t_{z'}^j)\cdot e^{it^j\cdot (\sigma_{(1,\cdot )}^j - \sigma^{j}_{(2,\cdot)})}dt^j\notag
\end{align}
Note that 
\begin{equation} 
\label{int2}
\begin{aligned}
&\int \Big(\sum_{\substack{z'}}\overline{s(z')} \partial_{t_{z'}^j}\Big)\Big(\sum_{\substack{z'}}s(z') \partial_{t_{z'}^j}\Big)\prod_{z'}p(t_{z'}^j)\cdot e^{it^j \cdot (\sigma_{(1,\cdot )}^j - \sigma^{j}_{(2,\cdot)})}dt^j\\
=&\int\Big(\sum_{z_1'\neq z_2'}\overline{s(z_1')}s(z_2')\frac{p'(t_{z_1'}^j)}{p(t_{z_1'}^j)}\frac{p'(t_{z_2'}^j)}{p(t_{z_2'}^j)}+\sum_{\substack{z'}}|s(z')|^2\frac{p''(t_{z'}^j)}{p(t_{z'}^j)}\Big) \prod_{z'}p(t_{z'}^j)\cdot e^{it^j \cdot (\sigma_{(1,\cdot )}^j - \sigma^{j}_{(2,\cdot)})}dt^j\\
=&\int\Bigg[\Big|\sum_{\substack{z'}}s(z')\frac{p'(t_{z'}^j)}{p(t_{z'}^j)}\Big|^2+\sum_{\substack{z'}}|s(z')|^2\Big(\frac{p''(t_{z'}^j)}{p(t_{z'}^j)}-(\frac{p'(t_{z'}^j)}{p(t_{z'}^j)})^2\Big)\Bigg]\prod_{z'}p(t_{z'}^j)\cdot e^{it^j \cdot (\sigma_{(1,\cdot )}^j - \sigma^{j}_{(2,\cdot)})}dt^j,
\end{aligned} 
\end{equation}
where 
\begin{equation}
\label{cin}
\Big|\sum_{\substack{z'}}s(z')\frac{p'(t_{z'}^j)}{p(t_{z'}^j)}\Big|^2\geq 0, \indent \sum\limits_{\substack{z'}}|s(z')|^2\Big(\frac{p''(t_{z'}^j)}{p(t_{z'}^j)}-(\frac{p'(t_{z'}^j)}{p(t_{z'}^j)})^2\Big)\geq \sum\limits_{\substack{z'}}( -M)=-L^{d-1}M . 
\end{equation}
Since $L^{d-1}M\int  \prod\limits_{z'}p(t_{z'}^j)\cdot e^{it^j \cdot (\sigma_{(1,\cdot )}^j - \sigma^{j}_{(2,\cdot)})}dt^j=L^{d-1}M\prod\limits_{z'}e^{-V\big(\sigma_{(1,z')}^j-\sigma_{(2,z')}^j\big)},$ 
  we basically have 
\begin{equation}
\label{Mff}
\mathbb{E}\Bigg[\Big|\sum\limits_{z'\in \mathbb{T}_L^{d-1}}s(z')\big[f(\sigma_{(1,z')})-f(\sigma_{(2,z')})\big]\Big|^2\Bigg]\leq L^{d-1}M,
\end{equation}
if we can just add the two inequalities (\ref{cin}) into the computation of the numerator of the expectation. To show this, we will need a giant reflection positivity on the whole torus. If $L$ is even, we let $\sigma_{\mathcal{L}}:\mathcal{L}\to \mathbb{R}^N$ and $\sigma_{\mathcal{R}}:\mathcal{R}\to \mathbb{R}^N$ be two symmetric parts of $\sigma$ (each has $\frac{L^d}{2}$ coordinates) and $\sigma_{(1,\cdot)}$ lies in $\sigma_{\mathcal{L}}$, $\sigma_{(2,\cdot)}$ lies in $\sigma_{\mathcal{R}}$ ($\mathcal{L}\cup\mathcal{R}=\mathbb{T}_L^d$). And we define $d\sigma_{\mathcal{L}}=\prod\limits_{z\in \mathcal{L}}d\sigma_z\cdot \prod\limits_{\substack{z\sim w\\z,w\in \mathcal{L}}}K(\sigma_z,\sigma_w)$, $d\sigma_{\mathcal{R}}=\prod\limits_{z\in \mathcal{R}}d\sigma_z\cdot \prod\limits_{\substack{z\sim w\\z,w\in \mathcal{R}}}K(\sigma_z,\sigma_w)$, we aim to prove that for some function $S:   (\mathbb{R}^N)^\mathcal{E}\to \mathbb{R}^{\geq 0}$, where $\mathcal{E}=\{v=(l_v,r_v):l_v\sim r_v, l_v\in \mathcal{L},r_v\in \mathcal{R}\}$,
\begin{equation}
\label{PDLL}
\int_{\sigma_\mathcal{L}}\int_{\sigma_\mathcal{R}}\int_t S(t)\prod_{\substack{v\in \mathcal{E}\\ q=1,\dots,N}}[e^{it_{v}^q(\sigma_{l_v}^q-\sigma_{r_v}^q)}p(t_v^q)]d\sigma_\mathcal{L}d\sigma_\mathcal{R}d_t\geq 0,
\end{equation}
and by choosing $S(t)$ to be the LHS minus RHS in (\ref{cin}), we already complete the proof. Now to prove (\ref{PDLL}), the technique is quite similar to (\ref{PDH}), we just note that
\begin{equation}
\begin{aligned}
&\int_{\sigma_\mathcal{L}}\int_{\sigma_\mathcal{R}}\int_t S(t)\prod_{\substack{v\in \mathcal{E}\\ q=1,\dots,N}}[e^{it_{v}^q(\sigma_{l_v}^q-\sigma_{r_v}^q)}p(t_v^q)]d\sigma_\mathcal{L}d\sigma_\mathcal{R}d_t\\
=&\int_t S(t)\Big [\int_{\sigma_\mathcal{L}}\int_{\sigma_\mathcal{R}} \prod_{\substack{v\in \mathcal{E}\\ q=1,\dots,N}}[e^{it_{v}^q(\sigma_{l_v}^q-\sigma_{r_v}^q)}p(t_v^q)]d\sigma_\mathcal{L}d\sigma_\mathcal{R}\Big]d_t\\
=&\int_t S(t)\Big [\int_{\sigma_\mathcal{L}} \prod_{\substack{v\in \mathcal{E}\\ q=1,\dots,N}}\Big[e^{it_{v}^q\sigma_{l_v}^q}\sqrt{p(t_v^q)}\Big]d\sigma_\mathcal{L}\Big]\Big[\int_{\sigma_\mathcal{R}} \prod_{\substack{v\in \mathcal{E}\\ q=1,\dots,N}}\Big[e^{-i t_v^q\sigma_{ r_v}^q}\sqrt{p(t_v^q)}\Big]d\sigma_\mathcal{R}\Big]d_t\\
=&\int_t S(t)\Bigg| \int_{\sigma_\mathcal{L}} \prod_{\substack{v\in \mathcal{E}\\ q=1,\dots,N}}\Big[e^{it_{v}^q\sigma_{l_v}^q}\sqrt{p(t_v^q)}\Big]d\sigma_\mathcal{L} \Bigg|^2 d_t\\
\geq & 0. 
\end{aligned} 
\end{equation}
We still need to check the absolute integrability. Note that $d\sigma_\mathcal{L}$ and $d\sigma_\mathcal{R}$ are both finite measures; we only need to check $\int_t|S(t)|\prod p(t_v^q)dt<\infty$. And it's enough to show that $\int_{t_v^q}|\frac{p'(t_v^q)}{p(t_v^q)}|^2\cdot p(t_v^q)dt_v^q<\infty$ and $\int_{t_v^q}\Big|\frac{p''(t_v^q)}{p(t_v^q)}-(\frac{p'(t_v^q)}{p(t_v^q)})^2+M\Big|\cdot p(t_v^q)dt_v^q<\infty$. In fact, 
$$\int|\frac{p'(t_v^q)}{p(t_v^q)}|^2\cdot p(t_v^q)dt_v^q=\int\frac{p''(t_v^q)}{p(t_v^q)}-\Big(\frac{p''(t_v^q)}{p(t_v^q)}-|\frac{p'(t_v^q)}{p(t_v^q)}|^2\Big)\cdot p(t_v^q)dt_v^q\leq \int\Big(|p''(t_v^q)|+Mp(t_v^q)\Big) dt_v^q<\infty,$$
and
$$\int\Big|\frac{p''(t_v^q)}{p(t_v^q)}-(\frac{p'(t_v^q)}{p(t_v^q)})^2+M\Big|\cdot p(t_v^q)dt_v^q\leq \int\Big(|p''(t_v^q)|+Mp(t_v^q)\Big) dt_v^q<\infty.$$
Here we use condition 3 in Definition \ref{conA} and $\int p(t_v^q)dt_v^q= e^{-V(0)}$. 

For $L$ odd, we play a small trick here: we fix the hyperplane $\sigma_{(\frac{L+3}{2},\cdot)}$ first; under this setting, $\sigma_\mathcal{L}$ and $\sigma_\mathcal{R}$ are chosen to be the two symmetric parts in the rest part (each has $\frac{L^{d-1}(L-1)}{2}$ coordinates). And we define $d\sigma_{\mathcal{L}}=\prod\limits_{z\in \mathcal{L}}d\sigma_z\cdot \prod\limits_{\substack{z\sim w\\z\in \mathcal{L},w\notin \mathcal{R}}}K(\sigma_z,\sigma_w)$, $d\sigma_{\mathcal{R}}=\prod\limits_{z\in \mathcal{R}}d\sigma_z\cdot \prod\limits_{\substack{z\sim w\\{z\in \mathcal{R},w\notin \mathcal{L}}}}K(\sigma_z,\sigma_w)$, $d\sigma_{(\frac{L+3}{2},\cdot)}=\prod\limits_{z\notin \mathcal{L}\cup \mathcal{R}}d\sigma_z\cdot \prod\limits_{\substack{z\sim w\\z,w\notin \mathcal{L}\cup \mathcal{R}}}K(\sigma_z,\sigma_w)$, in this sense we need to show that:
\begin{equation}
\begin{aligned}
&\int_{\sigma_{(\frac{L+3}{2},\cdot)}}\int_{\sigma_\mathcal{L}}\int_{\sigma_\mathcal{R}}\int_t S(t)\prod_{\substack{v\in \mathcal{E}\\ q=1,\dots,N}}[e^{it_{v}^q(\sigma_{l_v}^q-\sigma_{r_v}^q)}p(t_v^q)]d\sigma_\mathcal{L}d\sigma_\mathcal{R}d_td\sigma_{(\frac{L+3}{2},\cdot)}\\
=&\int_{\sigma_{(\frac{L+3}{2},\cdot)}}\int_t S(t)\Bigg| \int_{\sigma_\mathcal{L}} \prod_{\substack{v\in \mathcal{E}\\ q=1,\dots,N}}\Big[e^{it_{v}^q\sigma_{l_v}^q}\sqrt{p(t_v^q)}\Big]d\sigma_\mathcal{L} \Bigg|^2 d_td\sigma_{(\frac{L+3}{2},\cdot)}\\
\geq & 0. 
\end{aligned}
\end{equation}
Now, by Lemma \ref{PDL} and (\ref{Mff}) we have $\frac{1}{|\mathbb{T}_L^d|^2}\sum\limits_{z, w}\mathbb{E}_V\|\sigma_z^N - \sigma_w^N\|^2 \leq c(d)M$ and we simply need to take $C(N,d)=Nc(d)$.
\end{proof}
\medskip

Now we want to derive Theorem \ref{Ginz1} from this result. And we consider $\sigma: \mathbb{T}_L^d \to \mathbb{D}$ ($d \geq 3$), where $\mathbb{D} = \mathbb{R} \text{ or } \mathbb{Z}$, with Hamiltonian
\[
H(\sigma) = \sum_{z\sim w } V(\sigma_z  - \sigma_w ),
\]
and probability measure
\[
\mathbb{P}_\mathbb{D}(d\sigma)\propto \exp\big(-H(\sigma)\big)\delta_{\sigma_{z_0}=0}\cdot d\sigma,
\]
where $z_0$ is a fixed vertex in the torus, $d\sigma=\prod\limits_z d\sigma_z$ and $d\sigma_z$ is Lebesgue measure if $\mathbb{D}=\mathbb{R}$; counting measure $\sum\limits_{n\in \mathbb{Z}}\delta_n$ if $\mathbb{D}=\mathbb{Z}$. 

Again, we prove Theorem \ref{Ginz1} after we rewrite the statement of the theorem. We want to mention that this proof is based on a simple strategy: for any $B>0$, we use Theorem \ref{INF} to prove the result on the truncated model, where $\mathbb{D}$ is replaced by $\mathbb{D} \cap [-B, B)$, and take $B\to\infty$.
\begin{thm}
\label{Ginz}
If the potential $V$ satisfies condition $A$, then there exists a constant $C_1(d)>0$ that only depends on $d$, such that
\begin{equation}
\frac{1}{|\mathbb{T}_L^d|^2}\sum\limits_{z, w}\mathbb{E_D}(\sigma_z - \sigma_w)^2 \leq C_1(d)M ,
\end{equation}
\end{thm}
\begin{proof}
We first show the existence of this model, meaning that $ \int\exp\big(-H(\sigma)\big)\delta_{\sigma_{z_0}=0}\cdot d\sigma<\infty$.  We play a trick: let $\mathbb{T}_L^d=\{z_0,z_1,\dots,z_{|\mathbb{T}_L^d|-1}\}$, and $z_j\sim z_{j+1}$. An induction on $d$ will show that this is possible: for $d=1$ it is clear; for $d>2$ we first find a path on one hyperplane, and we move to the next, until we cover the whole torus. Now note that $e^{-V(x)}=\int p(t)\, e^{itx}dt\leq \int p(t)dt=  e^{-V(0)}$ is bounded, we only need to show that  
$$ \int
\prod_{i=1}^{|\mathbb{T}_L^d|-1}e^{-V(\sigma_{z_{i-1}}-\sigma_{z_i})} d\sigma_{z_i}<\infty.$$
Here $\sigma_{z_0}=0$. We can also prove it by induction: by condition 3 in Definition \ref{conA}, we have $\int e^{-V(\sigma_{z_1})}d\sigma_{z_1}<\infty$; also, we notice that
\begin{equation}
 \label{IND1}
 \int
\prod_{i=1}^{k}e^{-V(\sigma_{z_{i-1}}-\sigma_{z_i})} d\sigma_{z_i}=\int\Big[\prod_{i=1}^{k-1}e^{-V(\sigma_{z_{i-1}}-\sigma_{z_i})}d\sigma_{z_i}\Big]\cdot \Big[\int_{\sigma_{z_k}} e^{-V(\sigma_{z_{k-1}}-\sigma_{z_k})}d\sigma_{z_k} \Big],
\end{equation}
where $\Big[\int_{\sigma_k} e^{-V(\sigma_{z_{k-1}}-\sigma_{z_k})}d\sigma_k \Big]$ is a constant that does not depend on $\sigma_{z_{k-1}}$.

We note that the Hamiltonian and $d\sigma_z$ are unchanged after a global shift of $\sigma$, meaning that for any constant $c_0\in \mathbb{D}$, $\exp\big(-H(\sigma')\big)d\sigma'=\exp\big(-H(\sigma)\big)  d\sigma$ if $\sigma'_z=\sigma_z+c_0$ for any $z\in \mathbb{T}_L^d$. Consider a random variable $\gamma_0:=\max\limits_z \sigma_z -\min\limits_z \sigma_z$, denote $\frac{1}{|\mathbb{T}_L^d|^2}\sum\limits_{z, w}(\sigma_z - \sigma_w)^2$ by $\gamma$. Fix $B>0$ and $B\in \mathbb{D}$. For another model $\sigma^B:\mathbb{T}^d_L \to \mathbb{D}$ with the same Hamiltonian but $d\sigma^B_z$ is Lebesgue measure (or counting measure) on $\mathbb{D} \cap [-B, B)$,  and
\[
\mathbb{P}_\mathbb{D}^B(d\sigma^B) \propto \exp( -H(\sigma^B))d\sigma^B,
\]
if one similarly define  $\gamma^B_0:=\max\limits_z \sigma^B_z -\min\limits_z \sigma^B_z$, and $\gamma^B:=\frac{1}{|\mathbb{T}_L^d|^2}\sum\limits_{z, w}(\sigma^B_z - \sigma^B_w)^2$, by the translation invariance of the model, one may find out that $\mathbb{P}_\mathbb{D}[\gamma|\gamma_0]=\mathbb{P}^B_\mathbb{D}[\gamma^B|\gamma^B_0]$. Note also that $d\mathbb{P}_\mathbb{D}^B(\gamma^B,\gamma^B_0)\propto(2B-\gamma_0)^+\cdot d\mathbb{P}_\mathbb{D}(\gamma,\gamma_0)$, thus if we apply Theorem \ref{INF} on the truncated model $d\sigma^B$, we have that
\begin{equation}
\int \gamma^Bd\mathbb{P}_\mathbb{D}^B \leq C_1(d)M \int  d\mathbb{P}^B_\mathbb{D}, 
\end{equation}
which means that
\begin{equation}
\int \gamma (2B-\gamma_0)^+d\mathbb{P_D} \leq C_1(d)M \int  (2B-\gamma_0)^+ d\mathbb{P_D},
\end{equation}
thus,
$$\int \gamma {(1-\frac{\gamma_0}{2B})^+}{}d\mathbb{P}_\mathbb{D} \leq C_1(d)M  \int(1-\frac{\gamma_0}{2B})^+ \cdot d\mathbb{P}_\mathbb{D}\leq C_1(d)M,
$$
and we only need to show that $\int \gamma d\mathbb{P_D}<\infty$, so that the dominated convergence theorem can be used and we take $B\to +\infty$ to get the localization $\int \gamma d\mathbb{P_D}<C_1(d)M$.

For this, we basically need to prove that $\mathbb{E_D}(\sigma_z-\sigma_w)^2<\infty$ for any $z,w\in \mathbb{T}_L^d$. And it turns out that we only need to prove it for $z\sim w$, as we can always write $\sigma_z-\sigma_w$ as a finite sum of the differences between neighbours and that $(x_1+x_2+\dots+x_n)^2\leq n(x_1^2+\dots+x_n^2)$. For example, taking $z_1\sim z_0$, we aim to show that $\mathbb{E_D}\sigma_{z_1}^2<\infty$. 

The proof is very similar to the existence of the model. Since $e^{-V}$is bounded, our purpose is to prove that
$$ \int
\sigma_{z_1}^2\prod_{i=1}^{|\mathbb{T}_L^d|-1}e^{-V(\sigma_{z_{i-1}}-\sigma_{z_i})} d\sigma_{z_i}<\infty.$$
As in (\ref{IND1}), we have
\begin{equation}
 \int\sigma_{z_1}^2
\prod_{i=1}^{k}e^{-V(\sigma_{z_{i-1}}-\sigma_{z_i})} d\sigma_{z_i}=\int\sigma_{z_1}^2\Big[\prod_{i=1}^{k-1}e^{-V(\sigma_{z_{i-1}}-\sigma_{z_i})}d\sigma_{z_i}\Big]\cdot \Big[\int_{\sigma_{z_k}} e^{-V(\sigma_{z_{k-1}}-\sigma_{z_k})}d\sigma_{z_k} \Big],
\end{equation}
so that we can also do an induction on $k$. When $k=1$, the existence of the second moment of $e^{-V}$ comes from the existence of $p''$.
\end{proof}

\section{Long-range Order of the Villain Model}
\label{APFT}
Recall that the Villain model on a $d$-dimensional torus ($d\geq 3$) is a random configuration $\theta:\mathbb{T }_L^d \to [-\pi,\pi)$ with probability measure:
$$
\mathbb{P}_{Villain}(d\theta)\propto \prod_{z\sim w}\Big[\sum_{m\in\mathbb{Z}}\exp\Big(-\frac{\beta}{2}(\theta_z-\theta_w+2\pi m)^2\Big)\Big]d\theta,
$$ 
where  $\beta>0$ is the inverse temperature, $d\theta=\prod d\theta_z$ and $d\theta_z$ is the Lebesgue measure on $[-\pi,\pi)$. For convenience, we write $\sigma_z=(\cos\theta_z,\sin \theta_z)$. We will offer another proof for the long-range order of the Villain model on the torus of dimension $d\geq 3$ in the low-temperature regime in this section.

We restate the theorem and present the proof below. The proof idea is to take $f(\theta_z)=e^{i\theta_z}$, so that 
\begin{equation}
2-2\cdot\frac{1}{|\mathbb{T}_L^d|^2}\sum\limits_{z, w}\mathbb{E}_{Villain}\langle\sigma_z, \sigma_w\rangle=\frac{1}{|\mathbb{T}_L^d|^2}\sum\limits_{z, w}\mathbb{E}_{Villain}\|\sigma_z - \sigma_w\|^2 =\frac{1}{|\mathbb{T}_L^d|^2}\sum\limits_{z, w}\mathbb{E}_{Villain}|f(\theta_z) - f(\theta_w)|^2,
\end{equation}
and by Lemma \ref{PDL} we only need to find a bound for $M_f$. Parallel to the proof in Section \ref{SGA}, we apply a reflection positivity statement with some non-negative terms replaced by non-negative matrices. The computations in this section are for matrix reshaping and finding the non-negative blocks.
\begin{thm}
 We have long-range order: there exist constants $C_{Villain}>0$ and $\beta_M(d, C_{Villain})>0$ such that for all $\beta>\beta_M$  and $L \geq 2$, 
\begin{equation}
\label{VVVV}
\frac{1}{|\mathbb{T}_L^d|^2}\sum_{z, w}\mathbb{E}_{Villain}\langle\sigma_z,\sigma_w\rangle \geq C_{Villain}.
\end{equation}
\end{thm}
\begin{proof}
Note that $2-2\langle \sigma_z,\sigma_w\rangle=\|\sigma_z-\sigma_w\|^2=|e^{i\theta_z}-e^{i\theta_w}|^2$, and it turns out that we only need to bound  $\frac{1}{|\mathbb{T}_L^d|^2}\sum\limits_{z, w}\mathbb{E}_{Villain}|e^{i\theta_z}-e^{i\theta_w}|^2$. We want to use Lemma \ref{PDL} with $X=[-\pi,\pi)$, $d\mu=d\theta_z$, $K(x,y)=\sum\limits_{m\in\mathbb{Z}}\exp\Big(-\frac{\beta}{2}(x-y+2\pi m)^2\Big)$ and $f(x)=e^{ix}$. It's clear that $X$ is a separable metric space, $\mu$ is finite, and $K$ is bounded, symmetric, and measurable. We now check (\ref{PD2}). In fact, by Poisson summation we have 
$$
K(x,y)=\sum\limits_{m\in\mathbb{Z}}\exp\Big(-\frac{\beta}{2}(x-y+2\pi m)^2\Big)=\frac{1}{\sqrt{2\pi\beta}}\sum_{k\in \mathbb{Z}}e^{-\frac{k^2}{2\beta}}\cdot e^{ik(x-y)}.
$$
Thus, similar to (\ref{PDH}), we have that
\begin{equation} 
\begin{aligned}
&\int_{X^{{L}^{d-1}}}\int_{X^{{L}^{d-1}}} g(x )\overline{g(y )}\prod_{j}[K(x_j,y_j)d\mu(x_j)d\mu(y_j)]\\
=&\int_{X^{{L}^{d-1}}}\int_{X^{{L}^{d-1}}}g(x)\overline{g(y)}\prod_j \Big[\sum_{k_j}e^{ik_j(x_j-y_j)}p(k_j)\cdot d\mu(x_j)d\mu(y_j)\Big]\\
=&\sum_k \int_{X^{{L}^{d-1}}}\int_{X^{{L}^{d-1}}}g(x)\overline{g(y)}\prod_j \Big[e^{ik_j(x_j-y_j)}p(k_j)\cdot d\mu(x_j)d\mu(y_j)\Big] \\
=&\sum_k \Big(\prod_jp(k_j)\Big)\Big(\int_{X^{{L}^{d-1}}}g(x)\prod_j[e^{ik_j x_j}d\mu(x_j)]\Big)\cdot \Big(\int_{X^{{L}^{d-1}}}\overline{g(y)}\prod_j[e^{-ik_j y_j} d\mu(y_j)]\Big)\\
=&\sum_k \Big(\prod_jp(k_j)\Big) \Big|\int_{X^{{L}^{d-1}}}g(x)\prod_j[e^{ik_j x_j}d\mu(x_j)]\Big|^2 \\
\geq &0,
\end{aligned}
\end{equation}
where $p(k_j)=\frac{1}{\sqrt{2\pi\beta}}e^{-\frac{k_j^2}{2\beta}}$ is absolutely summable, $\mu$ is finite and $g$ is an $L^2$ function, so we can use Fubini's theorem.

It remains to compute $M_f=\frac{1}{L^{d-1}} \max\limits_{\substack{s(\cdot)\in\mathbb{S}^1 }}\mathbb{E}_{Villain}\Bigg[\Big|\sum\limits_{z'\in \mathbb{T}_L^{d-1}}s(z')\big[f(\theta_{(1,z')})-f(\theta_{(2,z')})\big]\Big|^2\Bigg]$.
Note that we basically want to establish a bound related to the following kernel:
\begin{align}
\label{UGUG}
&\left|\sum_{\substack{z'}}s(z') \left(e^{i\theta_{(1,z')}} - e^{i\theta_{(2,z')}}\right)\right|^2
\prod_{z'}K\big({\theta_{(1,z')}},{\theta_{(2,z')}}\big)\notag\\
=&\left|\sum_{\substack{z'}}s(z') \left(e^{i\theta_{(1,z')}} - e^{i\theta_{(2,z')}}\right)\right|^2 \prod_{z'}\Bigg[\frac{1}{\sqrt{2\pi\beta}}\sum_{k_{z'}}e^{-\frac{k_{z'}^2}{2\beta}}\cdot e^{ik_{z'}({\theta_{(1,z')}}-{\theta_{(2,z')}})}\Bigg]\notag\\
=&\Bigg[\sum_{\substack{z'_1,z'_2}}s(z'_1)\overline{s(z'_2)} \left(e^{i\theta_{(1,z'_1)}}\overline{ e^{i\theta_{(1,z'_2)}}}+e^{i\theta_{(2,z'_1)}}\overline{ e^{i\theta_{(2,z'_2)}}}-e^{i\theta_{(1,z'_1)}}\overline{ e^{i\theta_{(2,z'_2)}}}-e^{i\theta_{(2,z'_1)}}\overline{ e^{i\theta_{(1,z'_2)}}}\right)\Bigg]\cdot\\
&\prod_{z'}\Bigg[\frac{1}{\sqrt{2\pi\beta}}\sum_{k_{z'}}e^{-\frac{k_{z'}^2}{2\beta}}\cdot e^{ik_{z'}({\theta_{(1,z')}}-{\theta_{(2,z')}})}\Bigg].\notag
\end{align}
It has been written in a `matrix form' related to the orthogonal basis $\{\prod\limits_{z'}e^{ik_{z'}\theta_{(1,z')}}\}_k $ of $L^2(X^{{L}^{d-1}},\mu \otimes\dots\otimes\mu , \mathbb{C})$. The following equations will help us to reorganize this infinite matrix and find some positive definite component. 
\begin{equation}
\label{UG1}
\begin{aligned}
&e^{i\theta_{(1,z'_1)}}\overline{ e^{i\theta_{(1,z'_2)}}}\prod_{z'}\Bigg[\frac{1}{\sqrt{2\pi\beta}}\sum_{k_{z'}}e^{-\frac{k_{z'}^2}{2\beta}}\cdot e^{ik_{z'}({\theta_{(1,z')}}-{\theta_{(2,z')}})}\Bigg]\\
=&e^{i\theta_{(1,z'_1)}}\overline{ e^{i\theta_{(2,z'_2)}}}e^{i\theta_{(2,z'_2)}}\overline{ e^{i\theta_{(1,z'_2)}}}\prod_{z'}\Bigg[\frac{1}{\sqrt{2\pi\beta}}\sum_{k_{z'}}e^{-\frac{k_{z'}^2}{2\beta}}\cdot e^{ik_{z'}({\theta_{(1,z')}}-{\theta_{(2,z')}})}\Bigg]\\
=&e^{i\theta_{(1,z'_1)}}\overline{ e^{i\theta_{(2,z'_2)}}}\Bigg[\frac{1}{\sqrt{2\pi\beta}}\sum_{k_{z'_2}}e^{-\frac{k_{z'_2}^2}{2\beta}}\cdot e^{i(k_{z'_2}-1)({\theta_{(1,z_2')}}-{\theta_{(2,z_2')}})}\Bigg]\prod_{z'\neq z'_2}\Bigg[\frac{1}{\sqrt{2\pi\beta}}\sum_{k_{z'}}e^{-\frac{k_{z'}^2}{2\beta}}\cdot e^{ik_{z'}({\theta_{(1,z')}}-{\theta_{(2,z')}})}\Bigg]\\
=&e^{i\theta_{(1,z'_1)}}\overline{ e^{i\theta_{(2,z'_2)}}}\Bigg[\frac{1}{\sqrt{2\pi\beta}}\sum_{k_{z'_2}}e^{-\frac{(k_{z'_2}+1)^2}{2\beta}}\cdot e^{ik_{z'_2}({\theta_{(1,z_2')}}-{\theta_{(2,z_2')}})}\Bigg]\prod_{z'\neq z'_2}\Bigg[\frac{1}{\sqrt{2\pi\beta}}\sum_{k_{z'}}e^{-\frac{k_{z'}^2}{2\beta}}\cdot e^{ik_{z'}({\theta_{(1,z')}}-{\theta_{(2,z')}})}\Bigg]\\
=&e^{i\theta_{(1,z'_1)}}\overline{ e^{i\theta_{(2,z'_2)}}}\sum_{k}\frac{b_{k_{z_2'}+1}}{b_{k_{z_2'}}}\prod_{z'}\Bigg[\frac{1}{\sqrt{2\pi\beta}}e^{-\frac{k_{z'}^2}{2\beta}}\cdot e^{ik_{z'}({\theta_{(1,z')}}-{\theta_{(2,z')}})}\Bigg],
\end{aligned}
\end{equation}
where we write $b_{n}:=e^{-\frac{n^2} {2\beta}}$ for convenience. Similarly,
\begin{equation}
\label{UG2}
\begin{aligned}
&e^{i\theta_{(2,z'_1)}}\overline{ e^{i\theta_{(2,z'_2)}}}\prod_{z'}\Bigg[\frac{1}{\sqrt{2\pi\beta}}\sum_{k_{z'}}e^{-\frac{k_{z'}^2}{2\beta}}\cdot e^{ik_{z'}({\theta_{(1,z')}}-{\theta_{(2,z')}})}\Bigg]\\
=&e^{i\theta_{(1,z'_1)}}\overline{ e^{i\theta_{(2,z'_2)}}}e^{i\theta_{(2,z'_1)}}\overline{ e^{i\theta_{(1,z'_1)}}}\prod_{z'}\Bigg[\frac{1}{\sqrt{2\pi\beta}}\sum_{k_{z'}}e^{-\frac{k_{z'}^2}{2\beta}}\cdot e^{ik_{z'}({\theta_{(1,z')}}-{\theta_{(2,z')}})}\Bigg]\\
=&e^{i\theta_{(1,z'_1)}}\overline{ e^{i\theta_{(2,z'_2)}}}\Bigg[\frac{1}{\sqrt{2\pi\beta}}\sum_{k_{z'_1}}e^{-\frac{k_{z'_1}^2}{2\beta}}\cdot e^{i(k_{z'_1}-1)({\theta_{(1,z_1')}}-{\theta_{(2,z_1')}})}\Bigg]\prod_{z'\neq z'_1}\Bigg[\frac{1}{\sqrt{2\pi\beta}}\sum_{k_{z'}}e^{-\frac{k_{z'}^2}{2\beta}}\cdot e^{ik_{z'}({\theta_{(1,z')}}-{\theta_{(2,z')}})}\Bigg]\\
=&e^{i\theta_{(1,z'_1)}}\overline{ e^{i\theta_{(2,z'_2)}}}\Bigg[\frac{1}{\sqrt{2\pi\beta}}\sum_{k_{z'_1}}e^{-\frac{(k_{z'_1}+1)^2}{2\beta}}\cdot e^{ik_{z'_1}({\theta_{(1,z_1')}}-{\theta_{(2,z_1')}})}\Bigg]\prod_{z'\neq z'_1}\Bigg[\frac{1}{\sqrt{2\pi\beta}}\sum_{k_{z'}}e^{-\frac{k_{z'}^2}{2\beta}}\cdot e^{ik_{z'}({\theta_{(1,z')}}-{\theta_{(2,z')}})}\Bigg]\\
=&e^{i\theta_{(1,z'_1)}}\overline{ e^{i\theta_{(2,z'_2)}}}\sum_{k}\frac{b_{k_{z_1'}+1}}{b_{k_{z_1'}}}\prod_{z'}\Bigg[\frac{1}{\sqrt{2\pi\beta}}e^{-\frac{k_{z'}^2}{2\beta}}\cdot e^{ik_{z'}({\theta_{(1,z')}}-{\theta_{(2,z')}})}\Bigg],
\end{aligned}
\end{equation}
and
\begin{flalign}
\label{UG3}
&e^{i\theta_{(2,z'_1)}}\overline{ e^{i\theta_{(1,z'_2)}}}\prod_{z'}\Bigg[\frac{1}{\sqrt{2\pi\beta}}\sum_{k_{z'}}e^{-\frac{k_{z'}^2}{2\beta}}\cdot e^{ik_{z'}({\theta_{(1,z')}}-{\theta_{(2,z')}})}\Bigg]\notag\\
\quad=&e^{i\theta_{(1,z'_1)}}\overline{ e^{i\theta_{(2,z'_2)}}}e^{i\theta_{(2,z'_1)}}e^{i\theta_{(2,z'_2)}}\overline{ e^{i\theta_{(1,z'_1)}}}\overline{ e^{i\theta_{(1,z'_2)}}}\prod_{z'}\Bigg[\frac{1}{\sqrt{2\pi\beta}}\sum_{k_{z'}}e^{-\frac{k_{z'}^2}{2\beta}}\cdot e^{ik_{z'}({\theta_{(1,z')}}-{\theta_{(2,z')}})}\Bigg]\notag\\ 
\quad=&e^{i\theta_{(1,z'_1)}}\overline{ e^{i\theta_{(2,z'_2)}}}\sum_{k}\Bigg[(\mathds{1}_{z_1'\neq z_2'})\frac{b_{k_{z_1'}+1}}{b_{k_{z_1'}}}\cdot \frac{b_{k_{z_2'}+1}}{b_{k_{z_2'}}}+(\mathds{1}_{z_1'= z_2'})\frac{b_{k_{z_1'}+2}}{b_{k_{z_1'}}}\Bigg]\cdot \prod_{z'}\Bigg[\frac{1}{\sqrt{2\pi\beta}}e^{-\frac{k_{z'}^2}{2\beta}}\cdot e^{ik_{z'}({\theta_{(1,z')}}-{\theta_{(2,z')}})}\Bigg].&
\end{flalign}
Note that in this proof, we can always interchange the order of summation due to the absolute integrability, which comes from $\sum\limits_{k_{z'}}e^{-\frac{k_{z'}^2}{2\beta}}<\infty$.
Now plugging (\ref{UG1}), (\ref{UG2}) and (\ref{UG3}) into (\ref{UGUG}), we have
\begin{equation}  
\label{UGB}
\begin{aligned}
&\left|\sum_{\substack{z'}}s(z') \left(e^{i\theta_{(1,z')}} - e^{i\theta_{(2,z')}}\right)\right|^2
\prod_{z'}K\big({\theta_{(1,z')}},{\theta_{(2,z')}}\big)\\
=&-\sum_k\Bigg[\sum_{z'_1,z'_2}s(z_1')\overline{s(z_2')}e^{i\theta_{(1,z'_1)}}\overline{ e^{i\theta_{(2,z'_2)}}}(\frac{b_{k_{z_1'}+1}}{b_{k_{z_1'}}}-1)(\frac{b_{k_{z_2'}+1}}{b_{k_{z_2'}}}-1)\Bigg]\prod_{z'}\Bigg[\frac{1}{\sqrt{2\pi\beta}}e^{-\frac{k_{z'}^2}{2\beta}}\cdot e^{ik_{z'}({\theta_{(1,z')}}-{\theta_{(2,z')}})}\Bigg]\\
&+\sum_k\Bigg[\sum_{z' }s(z')\overline{s(z')}e^{i\theta_{(1,z')}}\overline{ e^{i\theta_{(2,z')}}}\Big((\frac{b_{k_{z'}+1}}{b_{k_{z'}}})^2-\frac{b_{k_{z'}+2}}{b_{k_{z'}}}\Big)\Bigg]\prod_{z'}\Bigg[\frac{1}{\sqrt{2\pi\beta}}e^{-\frac{k_{z'}^2}{2\beta}}\cdot e^{ik_{z'}({\theta_{(1,z')}}-{\theta_{(2,z')}})}\Bigg]\\
=&-\sum_k \Big[\sum_{z'}s(z')e^{i\theta_{(1,z')}}(\frac{b_{k_{z'}+1}}{b_{k_{z'}}}-1)\Big]\overline{\Big[\sum_{z'}s(z')e^{i\theta_{(2,z')}}(\frac{b_{k_{z'}+1}}{b_{k_{z'}}}-1)\Big]}\prod_{z'}\Bigg[\frac{1}{\sqrt{2\pi\beta}}e^{-\frac{k_{z'}^2}{2\beta}}\cdot e^{ik_{z'}({\theta_{(1,z')}}-{\theta_{(2,z')}})}\Bigg]\\
&+\sum_k\Bigg[\sum_{z' }e^{i\theta_{(1,z')}}\overline{ e^{i\theta_{(2,z')}}}\Big(\frac{b_{k_{z'}+1}^2-b_{k_{z'}+2}b_{k_{z'}}}{b_{k_{z'}}^2}\Big)\Bigg]\prod_{z'}\Bigg[\frac{1}{\sqrt{2\pi\beta}}e^{-\frac{k_{z'}^2}{2\beta}}\cdot e^{ik_{z'}({\theta_{(1,z')}}-{\theta_{(2,z')}})}\Bigg].
\end{aligned}
\end{equation}
We first consider the contribution of the first part of this summation to the integral in the numerator of $M_f$ and show that it's non-positive. In fact, it is already written in the form of the sum of non-positive definite submatrices. To rigorously show it, for $L$ even, we again let $\theta_{\mathcal{L}}:\mathcal{L}\to [-\pi,\pi)$ and $\theta_{\mathcal{R}}:\mathcal{R}\to [-\pi,\pi)$ be two symmetric parts of $\theta$ (each has $\frac{L^d}{2}$ coordinates) and $\theta_{(1,\cdot)}$ lies in $\theta_{\mathcal{L}}$, $\theta_{(2,\cdot)}$ lies in $\theta_{\mathcal{R}}$($\mathcal{L}\cup\mathcal{R}=\mathbb{T}_L^d$). And we define $d\theta_{\mathcal{L}}=\prod\limits_{z\in \mathcal{L}}d\theta_z\cdot \prod\limits_{\substack{z\sim w\\z,w\in \mathcal{L}}}K(\theta_z,\theta_w)$, $d\theta_{\mathcal{R}}=\prod\limits_{z\in \mathcal{R}}d\theta_z\cdot \prod\limits_{\substack{z\sim w\\z,w\in \mathcal{R}}}K(\theta_z,\theta_w)$, $\mathcal{E}=\{v=(l_v,r_v):l_v\sim r_v, l_v\in \mathcal{L},r_v\in \mathcal{R}\}$, thus we have
\begin{equation}
\begin{aligned}
&\int_{\theta_\mathcal{L}}\int_{\theta_\mathcal{R}}\sum_k \Big[\sum_{z'}s(z')e^{i\theta_{(1,z')}}(\frac{b_{k_{z'}+1}}{b_{k_{z'}}}-1)\Big]\overline{\Big[\sum_{z'}s(z')e^{i\theta_{(2,z')}}(\frac{b_{k_{z'}+1}}{b_{k_{z'}}}-1)\Big]}\prod_{v\in\mathcal{E}}\Bigg[\frac{1}{\sqrt{2\pi\beta}}e^{-\frac{k_{v}^2}{2\beta}}\cdot e^{ik_{v}({\theta_{l_v}}-{\theta_{r_v}})}\Bigg]d\theta_\mathcal{L}d\theta_\mathcal{R}\\
=&\sum_k\Bigg|\int_{\theta_\mathcal{L}}\Big[\sum_{z'}s(z')e^{i\theta_{(1,z')}}(\frac{b_{k_{z'}+1}}{b_{k_{z'}}}-1)\Big]\prod_{v\in\mathcal{E}}\Bigg[\sqrt{\frac{1}{\sqrt{2\pi\beta}}e^{-\frac{k_{v}^2}{2\beta}}}\cdot e^{ik_{v}{\theta_{l_v}}}\Bigg]d\theta_\mathcal{L}\Bigg|^2\\
\geq &0.
\end{aligned}
\end{equation}
For $L$ odd, it's similar to Theorem \ref{INF}; we omit it here. Now we work on the second part in the summation of (\ref{UGB}). In fact,

\begin{align*}
&\sum_k\Bigg[\sum_{z' }e^{i\theta_{(1,z')}}\overline{ e^{i\theta_{(2,z')}}}\Big(\frac{b_{k_{z'}+1}^2-b_{k_{z'}+2}b_{k_{z'}}}{b_{k_{z'}}^2}\Big)\Bigg]\prod_{z'}\Bigg[\frac{1}{\sqrt{2\pi\beta}}e^{-\frac{k_{z'}^2}{2\beta}}\cdot e^{ik_{z'}({\theta_{(1,z')}}-{\theta_{(2,z')}})}\Bigg]\\
=&\sum_k\Bigg[\sum_{z'_1 }e^{i\theta_{(1,z_1')}}\overline{ e^{i\theta_{(2,z'_1)}}}\Big(\frac{b_{k_{z'_1}+2}b_{k_{z_1'}}}{b_{k_{z_1'}}^2}(e^{\frac{1}{\beta}}-1)\Big)\Bigg]\prod_{z'}\Bigg[\frac{1}{\sqrt{2\pi\beta}}e^{-\frac{k_{z'}^2}{2\beta}}\cdot e^{ik_{z'}({\theta_{(1,z')}}-{\theta_{(2,z')}})}\Bigg]\\
=&(e^{\frac{1}{\beta}}-1)\sum_k\Bigg[\sum_{z'_1 }e^{i\theta_{(1,z_1')}}\overline{ e^{i\theta_{(2,z'_1)}}}\Big(\frac{b_{k_{z_1'}+2}}{b_{k_{z_1'}}}\Big)\frac{1}{\sqrt{2\pi\beta}}e^{-\frac{k_{z_1'}^2}{2\beta}}\cdot e^{ik_{z_1'}({\theta_{(1,z_1')}}-{\theta_{(2,z_1')}})}\prod_{z'\neq z'_1}\Big[\frac{1}{\sqrt{2\pi\beta}}e^{-\frac{k_{z'}^2}{2\beta}}\cdot e^{ik_{z'}({\theta_{(1,z')}}-{\theta_{(2,z')}})}\Big]\Bigg]\\
=&(e^{\frac{1}{\beta}}-1)\sum_k\Bigg[\sum_{z'_1 }\overline{e^{i\theta_{(1,z_1')}}}{ e^{i\theta_{(2,z'_1)}}}\frac{1}{\sqrt{2\pi\beta}}e^{-\frac{(k_{z_1'}+2)^2}{2\beta}}\cdot e^{i(k_{z_1'}+2)({\theta_{(1,z_1')}}-{\theta_{(2,z_1')}})}\prod_{z'\neq z'_1}\Big[\frac{1}{\sqrt{2\pi\beta}}e^{-\frac{k_{z'}^2}{2\beta}}\cdot e^{ik_{z'}({\theta_{(1,z')}}-{\theta_{(2,z')}})}\Big]\Bigg]\\
=&(e^{\frac{1}{\beta}}-1)\sum_k\Bigg[\sum_{z'_1 }\overline{e^{i\theta_{(1,z_1')}}}{ e^{i\theta_{(2,z'_1)}}}\Bigg]\prod_{z'}\Bigg[\frac{1}{\sqrt{2\pi\beta}}e^{-\frac{k_{z'}^2}{2\beta}}\cdot e^{ik_{z'}({\theta_{(1,z')}}-{\theta_{(2,z')}})}\Bigg]\\
=&(e^{\frac{1}{\beta}}-1)\Bigg[\sum_{z'_1 }\overline{e^{i\theta_{(1,z_1')}}}{ e^{i\theta_{(2,z'_1)}}}\Bigg]\prod_{z'}\Bigg[\frac{1}{\sqrt{2\pi\beta}}\sum_{k_{z'}}e^{-\frac{k_{z'}^2}{2\beta}}\cdot e^{ik_{z'}({\theta_{(1,z')}}-{\theta_{(2,z')}})}\Bigg]\\
=&(e^{\frac{1}{\beta}}-1)\Bigg[\sum_{z'_1 }\overline{e^{i\theta_{(1,z_1')}}}{ e^{i\theta_{(2,z'_1)}}}\Bigg]\prod_{z'}K\big({\theta_{(1,z')}},{\theta_{(2,z')}}\big).
\end{align*}
So the contribution of this term in the numerator of $\mathbb{E}_{Villain}\Bigg[\Big|\sum\limits_{z'\in \mathbb{T}_L^{d-1}}s(z')\big[f(\theta_{(1,z')})-f(\theta_{(2,z')})\big]\Big|^2\Bigg]$ is exactly $\int(e^{\frac{1}{\beta}}-1)\Big[\sum\limits_{z'_1 }\overline{e^{i\theta_{(1,z_1')}}}e^{i\theta_{(2,z_1')}}\Big]\prod\limits_{z\sim w}K(\theta_z,\theta_w)d\theta$, by the symmetry of $\theta \to -\theta$, it's equivalent to $\int(e^{\frac{1}{\beta}}-1)\Big[\sum\limits_{z'_1 }\cos(-\theta_{(1,z_1')}+\theta_{(2,z_1')})\Big]\prod\limits_{z\sim w}K(\theta_z,\theta_w)d\theta$, which is strictly smaller than $(e^{\frac{1}{\beta}}-1)L^{d-1}\int\prod\limits_{z\sim w}K(\theta_z,\theta_w)d\theta$, thus, we have $M_f\leq e^{\frac{1}{\beta}}-1$. As $\beta\to\infty$, this goes to 0, and we can pick suitable $C_{Villain}<1$ and $\beta_M(d,C_{Villain})$ such that
$$
\frac{1}{|\mathbb{T}_L^d|^2}\sum\limits_{z, w}\mathbb{E}_{Villain}|e^{i\theta_z}-e^{i\theta_w}|^2 \leq 2-2C_{Villain}
$$
for any $\beta>\beta_M$, and this is equivalent to (\ref{VVVV}).
\end{proof}
\section{Long-range Order of the Villain $O(N)$ Model}

\label{LOF}
Recall Definition \ref{Big} for the definition of the Villain $O(N)$ model. We will discuss the results related to the Villain $O(N)$ model and other rotation-invariant kernels on the sphere in this section. 

We first rewrite the statement of Theorem \ref{VN0} and present the proof. The proof strategy is a combination of Theorem \ref{INF0} and Theorem \ref{V0}. As in the proof of Theorem \ref{INF0}, we take $f_j(\sigma_z):=\sigma_z^j$ and try to find a bound for $M_{f_j}$. On the other hand, for the Villain model we managed to prove that as $\beta\to \infty$,
$$
M_{f}\leq \frac{1}{L^{d-1}}(e^{\frac{1}{\beta}}-1)\mathbb{E}_{Villain}\sum_{z_1'\in\mathbb{T}_L^{d-1}}\cos(-\theta_{(1,z_1')}+\theta_{(2,z_1')})\leq e^{\frac{1}{\beta}}-1\to 0,
$$
so for this Villain $O(N)$ model we seek a similar inequality and manage to prove that
$$
\sum_{j=1}^N M_{f_{j}}\leq \frac{1}{L^{d-1}}(e^{\frac{N-1}{\beta}}-1)\mathbb{E}_{Villain}^N\sum_{z_1'\in\mathbb{T}_L^{d-1}}\langle \sigma_{(1,z_1')},\sigma_{(2,z_1')} \rangle\leq e^{\frac{N-1}{\beta}}-1\to 0.
$$

\begin{thm}
\label{VN}
We have long-range order for Villain $O(N)$ model on the torus of dimension $d\geq 3$ in the low-temperature regime: for $N\geq 3$, there exist constants $C^N_{Villain}>0$ and $\beta_M^N(d,C^N_{Villain})>0$, such that for all $\beta>\beta_{M}^N$ and $L \geq 2$,
\begin{equation}
\label{VN01}
\frac{1}{|\mathbb{T}_L^d|^2}\sum_{z, w}\mathbb{E}^N_{Villain}\langle\sigma_z,\sigma_w\rangle \geq C_{Villain}^N.
\end{equation}
\end{thm}
\begin{proof}
Let $f_j(\sigma_z):=\sigma_z^j$. Again, we want to use Lemma \ref{PDL} with $X=\mathbb{S}^{N-1}$, $d\mu=d\sigma_z$, $K(x,y)=h_{\beta}(x,y)$ and $f=f_j$. It is clear that $X$ is a separable metric space, $\mu$ is a finite measure, $K$ is symmetric. Since the heat kernel is continuous, we also have that $K$ is bounded and measurable. Recall the following expansion of $h_\beta^{\mathbb{S}^{N-1}}$:
\begin{equation}
h_\beta^{\mathbb{S}^{N-1}}(x,y)=\sum_{l=0}^\infty e^{-\frac{l(l+N-2)}{2\beta}}\sum_{m=1}^{d_l}Y_{l,m}(x)\overline{Y_{l,m}(y)},
\end{equation}
where $d_l=\frac{(2l+N-2)(l+N-3)!}{l! (N-2)!}$, $Y_{l,m}$ are the spherical harmonics and form an orthonormal basis of $L^2(\mathbb{S}^{N-1},d\mu,\mathbb{C})$\cite{DaiXu}. Moreover, for any $l\geq0$, $Y_{l,m}$ form an orthonormal basis of $\mathcal{H}_l$, the space of spherical harmonics of degree $l$ on the sphere. So
$$
\int Y_{l,m}(x)\overline{Y_{l',m'}(x)}d\mu=\mathds{1}_{l=l'}\cdot\mathds{1}_{m=m'}. 
$$
There are several equivalent ways to define these spherical harmonics; Lemma \ref{SPS} in Appendix shows that there is a suitable definition such that 
\begin{equation}
\label{++--}
x^N\cdot Y_{l,m}(x)=a_{l,m}^{-}Y_{l-1,m}(x)+a_{l,m}^+Y_{l+1,m}(x),
\end{equation}
where $x=(x^1,\dots,x^N)\in \mathbb{S}^{N-1}$. $a^{+}_{l,m}$, $a^{-}_{l,m}$ are real constants that depend only on $l,m$, and we set $Y_{l,m}=0$ if $m>d_l$ or $l< 0$. This makes sure that the equations with $m>d_l$ or $l< 0$ will be trivial. In fact, we have
$$a_{l,m}^+=\int Y_{l,m}(x)\overline{Y_{l+1,m}(x)}x^Nd\mu=\int Y_{l+1,m}(x)\overline{Y_{l,m}(x)}x^Nd\mu=a_{l+1,m}^-.$$
Furthermore
$$
|a_{l,m}^+|^2=\Big|\int Y_{l,m}(x)\overline{Y_{l+1,m}(x)}x^Nd\mu\Big|^2\leq\int |Y_{l,m}(x)|^2 d\mu\int| \overline{Y_{l+1,m}(x)}x^N|^2d\mu\leq\int |Y_{l,m}(x)|^2 d\mu\int| \overline{Y_{l+1,m}(x)}|^2d\mu=1.
$$

Now we check (\ref{PD2}),
\begin{equation} 
\begin{aligned}
&\int_{X^{{L}^{d-1}}}\int_{X^{{L}^{d-1}}} g(x )\overline{g(y )}\prod_{k}[K(x_k,y_k)d\mu(x_k)d\mu(y_k)]\\
=&\int_{X^{{L}^{d-1}}}\int_{X^{{L}^{d-1}}}g(x)\overline{g(y)}\prod_k \Big[\sum_{l_k,m_k}Y_{l_k,m_k}(x_k)\overline{Y_{l_k,m_k}(y_k)}p(l_k)\cdot d\mu(x_k)d\mu(y_k)\Big]\\
=&\sum_{l,m} \int_{X^{{L}^{d-1}}}\int_{X^{{L}^{d-1}}}g(x)\overline{g(y)}\prod_k \Big[Y_{l_k,m_k}(x_k)\overline{Y_{l_k,m_k}(y_k)}p(l_k)\cdot d\mu(x_k)d\mu(y_k)\Big] \\
=&\sum_{l,m}\Big(\prod_k p(l_k)\Big)\cdot\Big(\int_{X^{{L}^{d-1}}}g(x)\prod_k[Y_{l_k,m_k}(x_k)d\mu(x_k)]\Big)\cdot \Big(\int_{X^{{L}^{d-1}}}\overline{g(y)}\prod_k[\overline{Y_{l_k,m_k}(y_k)} d\mu(y_k)]\Big)\\
=&\sum_{l,m}\Big(\prod_k p(l_k)\Big)\cdot\Big|\int_{X^{{L}^{d-1}}}g(x)\prod_k[Y_{l_k,m_k}(x_k)d\mu(x_k)]\Big|^2 \\
\geq &0,
\end{aligned}
\end{equation}
where $p(l_k)=e^{-\frac{l_k(l_k+N-2)}{2\beta}}$. To check the absolute integrability, we need the Cauchy-Schwarz inequality:
\begin{equation}
\label{CSF}
\begin{aligned}
&\sum_{l,m}\Big(\prod_k p(l_k)\Big)\cdot\Big(\int_{X^{{L}^{d-1}}}|g(x)|\prod_k[|Y_{l_k,m_k}(x_k)|d\mu(x_k)]\Big)\cdot \Big(\int_{X^{{L}^{d-1}}}|\overline{g(y)}|\prod_k[|\overline{Y_{l_k,m_k}(y_k)}|d\mu(y_k)]\Big)\\
\leq &\sum_{l,m}\Big(\prod_k p(l_k)\Big)\cdot\Big(\int_{X^{{L}^{d-1}}}|g(x)|^2 \prod_k[d\mu(x_k)]\Big)\cdot \Big(\int_{X^{{L}^{d-1}}} \prod_k[|Y_{l_k,m_k}(x_k)|^2d\mu(x_k)]\Big)\\
=&\sum_{l,m}\Big(\prod_k p(l_k)\Big)\cdot\Big(\int_{X^{{L}^{d-1}}}|g(x)|^2 \prod_k[d\mu(x_k)]\Big)\\
=&\prod_{k}\Big(\sum_{l_k}p(l_k)d_{l_k}\Big)\cdot\Big(\int_{X^{{L}^{d-1}}}|g(x)|^2 \prod_k[d\mu(x_k)]\Big)\\
<&\infty.
\end{aligned}    
\end{equation}
The last inequality comes from the fact that $g$ is a $L^2$ function and $\sum\limits_{l_k}d_{l_k}\cdot p(l_k)<\infty$.

And we want to find an upper bound for $M_{f_j}=\frac{1}{L^{d-1}} \max\limits_{\substack{s(\cdot)\in\mathbb{S}^1 }}\mathbb{E}_{Villain}^N\Bigg[\Big|\sum\limits_{z'\in \mathbb{T}_L^{d-1}}s(z')\big[\sigma^j_{(1,z')}-\sigma^j_{(2,z')})\big]\Big|^2\Bigg]$. As in the Villain model, we aim to work on the matrix representation of it on the basis of spherical harmonics. We first work with $j=N$. In fact,
\begin{equation} 
\label{+-UG}
\begin{aligned}
&\left|\sum_{\substack{z'}}s(z') \left(\sigma_{(1,z')}^N - \sigma_{(2,z')}^N\right)\right|^2
\prod_{z'}h_\beta^{\mathbb{S}^{N-1}}(\sigma_{(1,z')},\sigma_{(2,z')})\\
=&\Bigg(\sum_{\substack{z'}}s(z') \left(\sigma_{(1,z')}^N - \sigma_{(2,z')}^N\right)\Bigg)\Bigg(\sum_{\substack{z'}}\overline{s(z')} \left(\sigma_{(1,z')}^N - \sigma_{(2,z')}^N\right)\Bigg)\prod_{z'}\Bigg[\sum_{l_{z'},m_{z'}}e^{-\frac{l_{z'}(l_{z'}+N-2)}{2\beta}}Y_{l_{z'},m_{z'}}(\sigma_{(1,z')})\overline{Y_{l_{z'},m_{z'}}(\sigma_{(2,z')})}\Bigg]\\
=&\sum_{l,m}\Bigg(\sum_{\substack{z'}}s(z') \left(\sigma_{(1,z')}^N - \sigma_{(2,z')}^N\right)\Bigg)\Bigg(\sum_{\substack{z'}}\overline{s(z')} \left(\sigma_{(1,z')}^N - \sigma_{(2,z')}^N\right)\Bigg)\prod_{z'}\Bigg[e^{-\frac{l_{z'}(l_{z'}+N-2)}{2\beta}}Y_{l_{z'},m_{z'}}(\sigma_{(1,z')})\overline{Y_{l_{z'},m_{z'}}(\sigma_{(2,z')})}\Bigg]\\
=&\sum_{l,m}\sum_{z_1',z_2'}\Bigg(s(z_1')\overline{s(z_2')}\left(\sigma_{(1,z_1')}^N - \sigma_{(2,z_1')}^N\right)  \left(\sigma_{(1,z_2')}^N - 
\sigma_{(2,z_2')}^N\right)\prod_{z'}\Big[e^{-\frac{l_{z'}(l_{z'}+N-2)}{2\beta}}Y_{l_{z'},m_{z'}}(\sigma_{(1,z')})\overline{Y_{l_{z'},m_{z'}}(\sigma_{(2,z')})}\Big]\Bigg).
\end{aligned}
\end{equation}
Taking $g\equiv1$ in (\ref{CSF}), we will have absolute integrability in this formula as $\Big|\sum\limits_{z'\in \mathbb{T}_L^{d-1}}s(z')\big[\sigma^N_{(1,z')}-\sigma^N_{(2,z')})\big]\Big|$ is bounded. Now, for convenience, we define the operators $\mathscr{L}^{\pm}_z$ such that 
\[
\mathscr{L}_{(1,z_1')}^\pm \prod_{z'}\Big[Y_{l_{z'},m_{z'}}(\sigma_{(1,z')})\overline{Y_{l_{z'},m_{z'}}(\sigma_{(2,z')})}\Big]=Y_{l_{z_1'}\pm 1,m_{z_1'}}(\sigma_{(1,z_1')})\overline{Y_{l_{z_1'},m_{z_1'}}(\sigma_{(2,z_1')})}\prod_{z'\neq z_1'}\Big[Y_{l_{z'},m_{z'}}(\sigma_{(1,z')})\overline{Y_{l_{z'},m_{z'}}(\sigma_{(2,z')})}\Big],
\]
\begin{equation}
\label{MCL}
\mathscr{L}_{(2,z_1')}^\pm \prod_{z'}\Big[Y_{l_{z'},m_{z'}}(\sigma_{(1,z')})\overline{Y_{l_{z'},m_{z'}}(\sigma_{(2,z')})}\Big]=Y_{l_{z_1'},m_{z_1'}}(\sigma_{(1,z_1')})\overline{Y_{l_{z_1'}\pm 1,m_{z_1'}}(\sigma_{(2,z_1')})}\prod_{z'\neq z_1'}\Big[Y_{l_{z'},m_{z'}}(\sigma_{(1,z')})\overline{Y_{l_{z'},m_{z'}}(\sigma_{(2,z')})}\Big].
\end{equation}
These shift operators can also be seen as linear operators, as we can similarly define them on all bases of the space $L^2\Big((\mathbb{S}^{N-1})^{2L^{d-1}},\mathbb{C}\Big)$ in the form $\prod\limits_{z'}Y_{l^1_{z'},m^1_{z'}}(\sigma_{(1,z')})\prod\limits_{z'}\overline{Y_{l^2_{z'},m^2_{z'}}(\sigma_{(2,z')})}$. By (\ref{++--}), one may rewrite everything based on the same philosophy as in the Villain model:
\begin{equation}
\begin{aligned}
&\sum_{l,m}\sigma_{(1,z_1')}^N \sigma_{(2,z_2')}^N \prod_{z'}\Big[e^{-\frac{l_{z'}(l_{z'}+N-2)}{2\beta}}Y_{l_{z'},m_{z'}}(\sigma_{(1,z')})\overline{Y_{l_{z'},m_{z'}}(\sigma_{(2,z')})}\Big]\\
=&\sum_{l,m}\Big(a^+_{l_{z_1'},m_{z_1'}}\mathscr{L}_{(1,z_1')}^++a^-_{l_{z_1'},m_{z_1'}}\mathscr{L}_{(1,z_1')}^-\Big)\Big(a^+_{l_{z_2'},m_{z_2'}}\mathscr{L}_{(2,z_2')}^++a^-_{l_{z_2'},m_{z_2'}}\mathscr{L}_{(2,z_2')}^-\Big)\Phi(\sigma,l,m),\\
\end{aligned}
\end{equation}

\begin{equation}
\begin{aligned}
&\sum_{l,m}\sigma_{(1,z_1')}^N \sigma_{(1,z_2')}^N \prod_{z'}\Big[e^{-\frac{l_{z'}(l_{z'}+N-2)}{2\beta}}Y_{l_{z'},m_{z'}}(\sigma_{(1,z')})\overline{Y_{l_{z'},m_{z'}}(\sigma_{(2,z')})}\Big]\\
=&\sum_{l,m}\sigma_{(1,z_1')}^N\Big(a^+_{l_{z_2'},m_{z_2'}}\mathscr{L}_{(1,z_2')}^++a^-_{l_{z_2'},m_{z_2'}}\mathscr{L}_{(1,z_2')}^-\Big)\Phi(\sigma,l,m)\\
=&\sum_{l,m}\sigma_{(1,z_1')}^N\Big(a^+_{l_{z_2'}-1,m_{z_2'}}\frac{b_{l_{z'_2}-1}}{b_{l_{z'_2}}}\mathscr{L}_{(2,z_2')}^-+a^-_{l_{z_2'}+1,m_{z_2'}}\frac{b_{l_{z'_2}+1}}{b_{l_{z'_2}}}\mathscr{L}_{(2,z_2')}^+\Big)\Phi(\sigma,l,m),\\
=&\sum_{l,m}\Big(a^+_{l_{z_1'},m_{z_1'}}\mathscr{L}_{(1,z_1')}^++a^-_{l_{z_1'},m_{z_1'}}\mathscr{L}_{(1,z_1')}^-\Big)\Big(a^-_{l_{z_2'},m_{z_2'}}\frac{b_{l_{z'_2}-1}}{b_{l_{z'_2}}}\mathscr{L}_{(2,z_2')}^-+a^+_{l_{z_2'},m_{z_2'}}\frac{b_{l_{z'_2}+1}}{b_{l_{z'_2}}}\mathscr{L}_{(2,z_2')}^+\Big)\Phi(\sigma,l,m),\\
\end{aligned}
\end{equation}
and
\begin{equation}
\begin{aligned}
&\sum_{l,m}\sigma_{(2,z_1')}^N \sigma_{(2,z_2')}^N \prod_{z'}\Big[e^{-\frac{l_{z'}(l_{z'}+N-2)}{2\beta}}Y_{l_{z'},m_{z'}}(\sigma_{(1,z')})\overline{Y_{l_{z'},m_{z'}}(\sigma_{(2,z')})}\Big]\\
=&\sum_{l,m}\sigma_{(2,z_2')}^N\Big(a^+_{l_{z_1'},m_{z_1'}}\mathscr{L}_{(2,z_1')}^++a^-_{l_{z_1'},m_{z_1'}}\mathscr{L}_{(2,z_1')}^-\Big)\Phi(\sigma,l,m)\\
=&\sum_{l,m}\sigma_{(2,z_2')}^N\Big(a^+_{l_{z_1'}-1,m_{z_1'}}\frac{b_{l_{z'_1}-1}}{b_{l_{z'_1}}}\mathscr{L}_{(1,z_1')}^-+a^-_{l_{z_1'}+1,m_{z_1'}}\frac{b_{l_{z'_1}+1}}{b_{l_{z'_1}}}\mathscr{L}_{(1,z_1')}^+\Big)\Phi(\sigma,l,m),\\
=&\sum_{l,m}\Big(a^-_{l_{z_1'},m_{z_1'}}\frac{b_{l_{z'_1}-1}}{b_{l_{z'_1}}}\mathscr{L}_{(1,z_1')}^-+a^+_{l_{z_1'},m_{z_1'}}\frac{b_{l_{z'_1}+1}}{b_{l_{z'_1}}}\mathscr{L}_{(1,z_1')}^+\Big)\Big(a^+_{l_{z_2'},m_{z_2'}}\mathscr{L}_{(2,z_2')}^++a^-_{l_{z_2'},m_{z_2'}}\mathscr{L}_{(2,z_2')}^-\Big)\Phi(\sigma,l,m),\\
\end{aligned}
\end{equation}
where we write $\Phi(\sigma,l,m)=\prod\limits_{z'}\Big[e^{-\frac{l_{z'}(l_{z'}+N-2)}{2\beta}}Y_{l_{z'},m_{z'}} (\sigma_{(1,z')})\overline{Y_{l_{z'},m_{z'}}(\sigma_{(2,z')})}\Big]$ and $b_n=e^{-\frac{n(n+N-2)}{2\beta}}$ for short. Furthermore, for $z_1'\neq z_2'$ we have:
\begin{equation}
\begin{aligned}
&\sum_{l,m}\sigma_{(2,z_1')}^N \sigma_{(1,z_2')}^N \prod_{z'}\Big[e^{-\frac{l_{z'}(l_{z'}+N-2)}{2\beta}}Y_{l_{z'},m_{z'}}(\sigma_{(1,z')})\overline{Y_{l_{z'},m_{z'}}(\sigma_{(2,z')})}\Big]\\
=&\sum_{l,m}\Big(a^+_{l_{z_1'},m_{z_1'}}\mathscr{L}_{(2,z_1')}^++a^-_{l_{z_1'},m_{z_1'}}\mathscr{L}_{(2,z_1')}^-\Big)\Big(a^+_{l_{z_2'},m_{z_2'}}\mathscr{L}_{(1,z_2')}^++a^-_{l_{z_2'},m_{z_2'}}\mathscr{L}_{(1,z_2')}^-\Big)\Phi(\sigma,l,m)\\
=&\sum_{l,m}\Big(a^-_{l_{z_1'},m_{z_1'}}\frac{b_{l_{z'_1}-1}}{b_{l_{z'_1}}}\mathscr{L}_{(1,z_1')}^-+a^+_{l_{z_1'},m_{z_1'}}\frac{b_{l_{z'_1}+1}}{b_{l_{z'_1}}}\mathscr{L}_{(1,z_1')}^+\Big)\Big(a^-_{l_{z_2'},m_{z_2'}}\frac{b_{l_{z'_2}-1}}{b_{l_{z'_2}}}\mathscr{L}_{(2,z_2')}^-+a^+_{l_{z_2'},m_{z_2'}}\frac{b_{l_{z'_2}+1}}{b_{l_{z'_2}}}\mathscr{L}_{(2,z_2')}^+\Big)\Phi(\sigma,l,m),\\
\end{aligned}
\end{equation}
for $z_1'=z_2'$, it is simply
\begin{equation}
\label{p}
\begin{aligned}
&\sum_{l,m}\Big(a^+_{l_{z_1'},m_{z_1'}}\mathscr{L}_{(2,z_1')}^++a^-_{l_{z_1'},m_{z_1'}}\mathscr{L}_{(2,z_1')}^-\Big)\Big(a^+_{l_{z_1'},m_{z_1'}}\mathscr{L}_{(1,z_1')}^++a^-_{l_{z_1'},m_{z_1'}}\mathscr{L}_{(1,z_1')}^-\Big)\Phi(\sigma,l,m)\\
=&\sum_{l,m}\Big[\Big(a_{l_{z_1'},m_{z_1'}}^+\Big)^2\mathscr{L}_{(1,z_1')}^+\mathscr{L}_{(2,z_1')}^++\Big(a_{l_{z_1'},m_{z_1'}}^+a_{l_{z_1'},m_{z_1'}}^-\Big)\Big(\mathscr{L}_{(1,z_1')}^+\mathscr{L}_{(2,z_1')}^-+\mathscr{L}_{(1,z_1')}^-\mathscr{L}_{(2,z_1')}^+\Big)+\Big(a_{l_{z_1'},m_{z_1'}}^-\Big)^2\mathscr{L}_{(1,z_1')}^-\mathscr{L}_{(2,z_1')}^-\Big]\Phi(\sigma,l,m).
\end{aligned}
\end{equation}
And we can rewrite (\ref{+-UG}) into:
\begin{equation}
\label{LLUG}
\begin{aligned}
&-\sum_{l,m}\Bigg[\sum_{z_1'}s(z_1')\Big(a^-_{l_{z_1'},m_{z_1'}}\frac{b_{l_{z'_1}-1}}{b_{l_{z'_1}}}\mathscr{L}_{(1,z_1')}^-+a^+_{l_{z_1'},m_{z_1'}}\frac{b_{l_{z'_1}+1}}{b_{l_{z'_1}}}\mathscr{L}_{(1,z_1')}^+-a^+_{l_{z_1'},m_{z_1'}}\mathscr{L}_{(1,z_1')}^+-a^-_{l_{z_1'},m_{z_1'}}\mathscr{L}_{(1,z_1')}^-\Big)\Bigg]\cdot \\
&\Bigg[\sum_{z_2'}\overline{s(z_2')}\Big(a^-_{l_{z_2'},m_{z_2'}}\frac{b_{l_{z'_2}-1}}{b_{l_{z'_2}}}\mathscr{L}_{(2,z_2')}^-+a^+_{l_{z_2'},m_{z_2'}}\frac{b_{l_{z'_2}+1}}{b_{l_{z'_2}}}\mathscr{L}_{(2,z_2')}^+- a^+_{l_{z_2'},m_{z_2'}}\mathscr{L}_{(2,z_2')}^+-a^-_{l_{z_2'},m_{z_2'}}\mathscr{L}_{(2,z_2')}^-\Big)\Bigg]\Phi(\sigma,l,m)\\
&+\sum_{l,m}\sum_{z_1'}|s(z_1')|^2\Bigg[\Big[ -\Big(a^-_{l_{z_1'},m_{z_1'}}\Big)^2+\Big(\frac{b_{l_{z'_1}-1}}{b_{l_{z'_1}}}\Big)^2\Big(a^-_{l_{z_1'},m_{z_1'}}\Big)^2\Big]\mathscr{L}_{(1,z_1')}^-\mathscr{L}_{(2,z_1')}^-+\Big[-1+\frac{b_{l_{z_1'}-1}}{b_{l_{z_1'}}}\cdot \frac{b_{l_{z_1'}+1}}{b_{l_{z_1'}}}\Big]a^-_{l_{z_1'},m_{z_1'}}\\ & a^+_{l_{z_1'},m_{z_1'}}\Big[\mathscr{L}_{(1,z_1')}^-\mathscr{L}_{(2,z_1')}^+
+\mathscr{L}_{(1,z_1')}^+\mathscr{L}_{(2,z_1')}^-\Big]+\Big[-\Big(a^+_{l_{z_1'},m_{z_1'}}\Big)^2+\Big(\frac{b_{l_{z'_1}+1}}{b_{l_{z'_1}}}\Big)^2\Big(a^+_{l_{z_1'},m_{z_1'}}\Big)^2\Big]\mathscr{L}_{(1,z_1')}^+\mathscr{L}_{(2,z_1')}^+\Bigg]\Phi(\sigma,l,m).
\end{aligned}    
\end{equation}
The contribution of the first part of this summation in the numerator of $M_f$ should be non-positive. To show it, for $L$ even, we again let $\sigma_{\mathcal{L}}:\mathcal{L}\to \mathbb{S}^{N-1}$ and $\sigma_{\mathcal{R}}:\mathcal{R}\to \mathbb{S}^{N-1}$ be two symmetric parts of $\sigma$ (each has $\frac{L^d}{2}$ coordinates) and $\sigma_{(1,\cdot)}$ lies in $\sigma_{\mathcal{L}}$, $\sigma_{(2,\cdot)}$ lies in $\sigma_{\mathcal{R}}$ ($\mathcal{L}\cup\mathcal{R}=\mathbb{T}_L^d$). And we define $d\sigma_{\mathcal{L}}=\prod\limits_{z\in \mathcal{L}}d\sigma_z\cdot \prod\limits_{\substack{z\sim w\\z,w\in \mathcal{L}}}K(\sigma_z,\sigma_w)$, $d\sigma_{\mathcal{R}}=\prod\limits_{z\in \mathcal{R}}d\sigma_z\cdot \prod\limits_{\substack{z\sim w\\z,w\in \mathcal{R}}}K(\sigma_z,\sigma_w)$, $\mathcal{E}=\{v=(z^l_v,z^r_v):z^l_v\sim z^r_v, z^l_v\in \mathcal{L},z^r_v\in \mathcal{R}\}$. We have that 

\begin{equation}
\label{OE}
\begin{aligned}
&\int_{\sigma_\mathcal{L}}\int_{\sigma_\mathcal{R}}d\sigma_\mathcal{L}d\sigma_\mathcal{R}\cdot\sum_{l,m}\Bigg[\sum_{z_1'}s(z_1')\Big(a^-_{l_{z_1'},m_{z_1'}}\frac{b_{l_{z'_1}-1}}{b_{l_{z'_1}}}\mathscr{L}_{(1,z_1')}^-+a^+_{l_{z_1'},m_{z_1'}}\frac{b_{l_{z'_1}+1}}{b_{l_{z'_1}}}\mathscr{L}_{(1,z_1')}^+-a^+_{l_{z_1'},m_{z_1'}}\mathscr{L}_{(1,z_1')}^+-a^-_{l_{z_1'},m_{z_1'}}\mathscr{L}_{(1,z_1')}^-\Big)\Bigg]\cdot \\
&\Bigg[\sum_{z_2'}\overline{s(z_2')}\Big(a^-_{l_{z_2'},m_{z_2'}}\frac{b_{l_{z'_2}-1}}{b_{l_{z'_2}}}\mathscr{L}_{(2,z_2')}^-+a^+_{l_{z_2'},m_{z_2'}}\frac{b_{l_{z'_2}+1}}{b_{l_{z'_2}}}\mathscr{L}_{(2,z_2')}^+- a^+_{l_{z_2'},m_{z_2'}}\mathscr{L}_{(2,z_2')}^+-a^-_{l_{z_2'},m_{z_2'}}\mathscr{L}_{(2,z_2')}^-\Big)\Bigg]\prod_{v\in \mathcal{E}}\Bigg[b_{l_v}Y_{l_v,m_v}(\sigma_{z_v^l})\overline{Y_{l_v,m_v}(\sigma_{z_v^r})}\Bigg]\\
=&\sum_{l,m}\Bigg|\int_{\sigma_\mathcal{L}}d\sigma_\mathcal{L}\cdot\sum_{z_1'}s(z_1')\Big(a^-_{l_{z_1'},m_{z_1'}}\frac{b_{l_{z'_1}-1}}{b_{l_{z'_1}}}\mathscr{L}_{(1,z_1')}^-+a^+_{l_{z_1'},m_{z_1'}}\frac{b_{l_{z'_1}+1}}{b_{l_{z'_1}}}\mathscr{L}_{(1,z_1')}^+-a^+_{l_{z_1'},m_{z_1'}}\mathscr{L}_{(1,z_1')}^+-a^-_{l_{z_1'},m_{z_1'}}\mathscr{L}_{(1,z_1')}^-\Big)\\
&\prod_{v\in \mathcal{E}}\Big[\sqrt{b_{l_v}}Y_{l_{v},m_{v}}(\sigma_{z_v^l})\Big] \Bigg|^2\\
 \geq  &  0 , 
\end{aligned}    
\end{equation}
where we define $\mathscr{L}_{(1,z_1')}^\pm \prod\limits_{v\in \mathcal{E}}\Big[Y_{l_{v},m_{v}}(\sigma_{z_v^l})\Big]$ similarly to (\ref{MCL}) without conjugations. We use the same notation since they are indeed the same as linear operators on $L^2\Big((\mathbb{S}^{N-1})^{2L^{d-1}},\mathbb{C}\Big)$. For $L$ odd, the trick in Theorem \ref{INF} still works, and we omit the proof. The only difference is that $\mathcal{E}$ only consists of edges between $\sigma_{(1,\cdot)}$ and $\sigma_{(2,\cdot)}$, so we will need operators on  $L^2\Big((\mathbb{S}^{N-1})^{L^{d-1}},\mathbb{C}\Big)$ in (\ref{OE}). Since we interchange the order of $\int_{\sigma_\mathcal{L}}\int_{\sigma_\mathcal{R}}$ and $\sum\limits_{l,m}$, we still need to check the absolute integrability in (\ref{OE}). Note that $d\sigma_\mathcal{L}$ and $d\sigma_\mathcal{R}$ are finite measures, $|a_{l_{z_1'},m_{z_1'}}^\pm|\leq1$, we can again use (\ref{CSF}) with $g\equiv1$ and $\prod\limits_k p(l_k)$ replaced by $\prod\limits_v \Tilde{b_{l_v}}$, where $\Tilde{b_{l_v}}$ can be $b_{l_v}$ or $\frac{b_{l_v\pm 1}}{b_{l_v}}\cdot b_{l_v}$ or even $\Big(\frac{b_{l_v\pm 1}}{b_{l_v}}\Big)\Big(\frac{b_{l_v\pm 1}}{b_{l_v}}\Big)\cdot b_{l_v}$, since we have 16 different terms in (\ref{OE}) and when $z_v^l=(1,z_1')$ or $z_v^l=(1,z_2')$ the coefficient of $Y_{l_v,m_v}(\sigma_{z_v^l})\overline{Y_{l_v,m_v}(\sigma_{z_v^r})}$ will be more complicated. But it will always decay exponentially w.r.t $l_v$ and we have $\sum\limits_{l_v\geq 0} d_{l_v}\cdot\Tilde{b_{l_v}}< \infty$.

Now we work with the second part of (\ref{LLUG}). In fact,
it is 
\begin{equation}
\label{scrh}
\begin{aligned}
&\mathscr{H}^N_\beta(\sigma_{(1,\cdot)},\sigma_{(2,\cdot)}):=\sum_{l,m}\sum_{z_1'}\Bigg[\Big[ -\Big(a^-_{l_{z_1'},m_{z_1'}}\Big)^2+\Big(\frac{b_{l_{z'_1}-1}}{b_{l_{z'_1}}}\Big)^2\Big(a^-_{l_{z_1'},m_{z_1'}}\Big)^2\Big]\mathscr{L}_{(1,z_1')}^-\mathscr{L}_{(2,z_1')}^-+\Big[-1+\frac{b_{l_{z_1'}-1}}{b_{l_{z_1'}}}\cdot \frac{b_{l_{z_1'}+1}}{b_{l_{z_1'}}}\Big]a^-_{l_{z_1'},m_{z_1'}}\\ & a^+_{l_{z_1'},m_{z_1'}}\Big[\mathscr{L}_{(1,z_1')}^-\mathscr{L}_{(2,z_1')}^+
+\mathscr{L}_{(1,z_1')}^+\mathscr{L}_{(2,z_1')}^-\Big]+\Big[-\Big(a^+_{l_{z_1'},m_{z_1'}}\Big)^2+\Big(\frac{b_{l_{z'_1}+1}}{b_{l_{z'_1}}}\Big)^2\Big(a^+_{l_{z_1'},m_{z_1'}}\Big)^2\Big]\mathscr{L}_{(1,z_1')}^+\mathscr{L}_{(2,z_1')}^+\Bigg]\Phi(\sigma,l,m).\\
\end{aligned}
\end{equation}
This kernel looks very close to (\ref{p}), which is $\sigma^N_{(1,z')}\sigma_{(2,z')}^N\prod_{z'}h_\beta^{\mathbb{S}^{N-1}}(\sigma_{(1,z')},\sigma_{(2,z')})$. In fact, if we denote the projections from $L^2(\mathbb{S}^{N-1},d\mu,\mathbb{C})$ to $\mathcal{H}_{l_0}$ (the space of spherical harmonics of degree $l_0$) as $\mathcal{P}_{l_0}$ and the operator of multiplication by $x^j$ as $\mathcal{X}_j$, the integral kernel of $\mathcal{P}_{l_0-1}\mathcal{X}_N \mathcal{P}_{l_0}\mathcal{P}_{l_0}^*\mathcal{X}_N^* \mathcal{P}_{l_0-1}^*$ would be  
\begin{equation}
\label{KKKK}
\begin{aligned}
\mathcal{K}_{l_0,l_0-1}^N(x,y)&=\sum_{l_{z_1'},m_{z_1'}}\mathcal{P}_{l_0-1}\mathcal{X}_N \mathcal{P}_{l_0}Y_{l_{z_1'},m_{z_1'}}(x)\cdot \overline{\mathcal{P}_{l_0-1}\mathcal{X}_N \mathcal{P}_{l_0}Y_{l_{z_1'},m_{z_1'}}(y)}\\
&=\sum_{m_{z_1'}}\mathcal{P}_{l_0-1}\mathcal{X}_N Y_{l_0,m_{z_1'}}(x)\cdot \overline{\mathcal{P}_{l_0-1}\mathcal{X}_N Y_{l_0,m_{z_1'}}(y)}\\
&=\sum_{m_{z_1'}} \big(a_{l_0,m_{z_1'}}^-\big)^2Y_{l_0-1,m_{z_1'}}(x)\overline{Y_{l_0-1,m_{z_1'}}(y)}
. 
\end{aligned}
\end{equation}
We can similarly compute the integral kernel of $\mathcal{P}_{l_0+1}\mathcal{X}_N \mathcal{P}_{l_0}\mathcal{P}_{l_0}^*\mathcal{X}_N^* \mathcal{P}_{l_0+1}^*$:
\begin{equation}
\begin{aligned}
\mathcal{K}_{l_0,l_0+1}^N(x,y)&=\sum_{l_{z_1'},m_{z_1'}}\mathcal{P}_{l_0+1}\mathcal{X}_N \mathcal{P}_{l_0}Y_{l_{z_1'},m_{z_1'}}(x)\cdot \overline{\mathcal{P}_{l_0+1}\mathcal{X}_N \mathcal{P}_{l_0}Y_{l_{z_1'},m_{z_1'}}(y)}\\
&=\sum_{m_{z_1'}}\mathcal{P}_{l_0+1}\mathcal{X}_N Y_{l_0,m_{z_1'}}(x)\cdot \overline{\mathcal{P}_{l_0+1}\mathcal{X}_N Y_{l_0,m_{z_1'}}(y)}\\
&=\sum_{m_{z_1'}} \big(a_{l_0,m_{z_1'}}^+\big)^2Y_{l_0+1,m_{z_1'}}(x)\overline{Y_{l_0+1,m_{z_1'}}(y)},
\end{aligned}
\end{equation}
the integral kernel of $\mathcal{P}_{l_0-1}\mathcal{X}_N \mathcal{P}_{l_0}\mathcal{P}_{l_0}^*\mathcal{X}_N^* \mathcal{P}_{l_0+1}^*$:
\begin{equation}
\begin{aligned}
\mathcal{K}_{l_0-1,l_0+1}^N(x,y)&=\sum_{l_{z_1'},m_{z_1'}}\mathcal{P}_{l_0-1}\mathcal{X}_N \mathcal{P}_{l_0}Y_{l_{z_1'},m_{z_1'}}(x)\cdot \overline{\mathcal{P}_{l_0+1}\mathcal{X}_N \mathcal{P}_{l_0}Y_{l_{z_1'},m_{z_1'}}(y)}\\
&=\sum_{m_{z_1'}}\mathcal{P}_{l_0-1}\mathcal{X}_N Y_{l_0,m_{z_1'}}(x)\cdot \overline{\mathcal{P}_{l_0+1}\mathcal{X}_N Y_{l_0,m_{z_1'}}(y)}\\
&=\sum_{m_{z_1'}} \big(a_{l_0,m_{z_1'}}^-\big)\big(a_{l_0,m_{z_1'}}^+\big)Y_{l_0-1,m_{z_1'}}(x)\overline{Y_{l_0+1,m_{z_1'}}(y)},
\end{aligned}
\end{equation}
and the integral kernel of $\mathcal{P}_{l_0+1}\mathcal{X}_N \mathcal{P}_{l_0}\mathcal{P}_{l_0}^*\mathcal{X}_N^* \mathcal{P}_{l_0-1}^*$:
\begin{equation}
\begin{aligned}
\mathcal{K}_{l_0+1,l_0-1}^N(x,y)&=\sum_{l_{z_1'},m_{z_1'}}\mathcal{P}_{l_0+1}\mathcal{X}_N \mathcal{P}_{l_0}Y_{l_{z_1'},m_{z_1'}}(x)\cdot \overline{\mathcal{P}_{l_0-1}\mathcal{X}_N \mathcal{P}_{l_0}Y_{l_{z_1'},m_{z_1'}}(y)}\\
&=\sum_{m_{z_1'}}\mathcal{P}_{l_0+1}\mathcal{X}_N Y_{l_0,m_{z_1'}}(x)\cdot \overline{\mathcal{P}_{l_0-1}\mathcal{X}_N Y_{l_0,m_{z_1'}}(y)}\\
&=\sum_{m_{z_1'}} \big(a_{l_0,m_{z_1'}}^+\big)\big(a_{l_0,m_{z_1'}}^-\big)Y_{l_0+1,m_{z_1'}}(x)\overline{Y_{l_0-1,m_{z_1'}}(y)}
. 
\end{aligned}
\end{equation}
And we will have that
\begin{equation}
\begin{aligned}
&\mathscr{H}^N_\beta(\sigma_{(1,\cdot)},\sigma_{(2,\cdot)})=\sum_{z_1'}\sum_{l_{z_1'}}b_{l_{z_1'}}\Bigg(\Big[-1+\Big(\frac{b_{l_{z_1'}-1}}{b_{l_{z_1'}}}\Big)^2\Big]\mathcal{K}_{l_{z_1'},l_{z_1'}-1}^N(\sigma_{(1,z_1')}, \sigma_{(2,z_1')})+\Big[-1+\frac{b_{l_{z_1'}-1}}{b_{l_{z_1'}}}\cdot \frac{b_{l_{z_1'}+1}}{b_{l_{z_1'}}}\Big]\Big(\mathcal{K}_{l_{z_1'}+1,l_{z_1'}-1}^N\\
&+\mathcal{K}_{l_{z_1'}-1,l_{z_1'}+1}^N\Big)(\sigma_{(1,z_1')}, \sigma_{(2,z_1')})+\Big[-1+\Big(\frac{b_{l_{z_1'}+1}}{b_{l_{z_1'}}}\Big)^2\Big]\mathcal{K}_{l_{z_1'},l_{z_1'}+1}^N(\sigma_{(1,z_1')}, \sigma_{(2,z_1')})\Bigg)\prod_{z_2'\neq z_1'}h_\beta^{\mathbb{S}^{N-1}}(\sigma_{(1,z_2')},\sigma_{(2,z_2')}).
\end{aligned}    
\end{equation}
Now we define $\mathcal{K}^j_{l_1,l_2}$ similarly to $\mathcal{K}^N_{l_1,l_2}$, just with $\mathcal{X}_N$ and $\mathcal{X}_N^*$ replaced by $\mathcal{X}_j$ and $\mathcal{X}_j^*$. Note that, since the contribution of the first part of (\ref{LLUG}) is non-positive, we have already proved that

\begin{equation}
\label{,ldk}
\begin{aligned}
&Z_{Villain}^NM_{f_N}\leq \frac{1}{L^{d-1}}\int\sum_{z_1'}\sum_{l_{z_1'}}b_{l_{z_1'}}\Bigg(\Big[-1+\Big(\frac{b_{l_{z_1'}-1}}{b_{l_{z_1'}}}\Big)^2\Big]\mathcal{K}_{l_{z_1'},l_{z_1'}-1}^N(\sigma_{(1,z_1')}, \sigma_{(2,z_1')})+\Big[-1+\frac{b_{l_{z_1'}-1}}{b_{l_{z_1'}}}\cdot \frac{b_{l_{z_1'}+1}}{b_{l_{z_1'}}}\Big]\Big(\mathcal{K}_{l_{z_1'}+1,l_{z_1'}-1}^N
\\
&+\mathcal{K}_{l_{z_1'}-1,l_{z_1'}+1}^N\Big)(\sigma_{(1,z_1')}, \sigma_{(2,z_1')})+\Big[-1+\Big(\frac{b_{l_{z_1'}+1}}{b_{l_{z_1'}}}\Big)^2\Big]\mathcal{K}_{l_{z_1'},l_{z_1'}+1}^N(\sigma_{(1,z_1')}, \sigma_{(2,z_1')})\Bigg)\prod_{\substack{z\sim w\\\{z,w\}\neq\{(1,z_1'),(2,z_1')\}}}h_\beta^{\mathbb{S}^{N-1}}(\sigma_z,\sigma_w)d\sigma.
\end{aligned}    
\end{equation}
Here $Z_{Villain}^N:=\int_\sigma \prod_{z\sim w} h_\beta^{\mathbb{S}^{N-1}}(\sigma_z,\sigma_w)d\sigma$ is the partition function, so we basically get a bound for the numerator of $M_{f_N}$. By symmetry, we can replace $N$ with $j$ in this equation, since the definitions of $M_{f_N}$, $\mathcal{K}$ and $h_\beta^{\mathbb{S}^{N-1}}$ do not rely on the choices of $Y_{l,m}$. ($Y_{l,m}$ are not symmetric for $j$, but $\mathcal{P}_{l_0}$, $M_{f_j}$, $\mathcal{X}_j$ and $h_{\beta}$ are all symmetric for $j$.) Note that by Lemma \ref{PDL} we have
$$
2-2\frac{1}{|\mathbb{T}_L^d|^2}\sum_{z,w}\mathbb{E}_{Villain}^N\langle\sigma_z,\sigma_w\rangle=\frac{1}{|\mathbb{T}_L^d|^2}\sum_{z,w}\mathbb{E}_{Villain}^N\|\sigma_z-\sigma_w\|^2=\sum_{j=1}^N\Big(\frac{1}{|\mathbb{T}_L^d|^2}\sum_{z,w}\mathbb{E}_{Villain}^N(\sigma_z^j-\sigma_w^j)^2\Big)\leq \sum_{j=1}^N c\cdot M_{f_j}.
$$
By (\ref{,ldk}), we basically need to compute $\sum\limits_{j=1}^N\mathcal{K}_{l_1,l_2}^j$. Note that the integral kernel of $\sum\limits_{j=1}^N\mathcal{X}_j\mathcal{P}_{l_0}\mathcal{P}_{l_0}^*\mathcal{X}_j^*$ is
\begin{align}
\label{expansion}
&\sum_{j=1}^{N}\sum_{l_{z_1'},m_{z_1'}}\mathcal{X}_j \mathcal{P}_{l_0}Y_{l_{z_1'},m_{z_1'}}(x)\cdot \overline{\mathcal{X}_j \mathcal{P}_{l_0}Y_{l_{z_1'},m_{z_1'}}(y)}\notag\\
=&\sum_{j=1}^{N}\sum_{m_{z_1'}}\mathcal{X}_j Y_{l_0,m_{z_1'}}(x)\cdot \overline{\mathcal{X}_j Y_{l_0,m_{z_1'}}(y)}\notag\\
=&\langle x, y\rangle\sum_{m_{z_1'}=1}^{d_{l_0}} Y_{l_0,m_{z_1'}}(x)\overline{Y_{l_0,m_{z_1'}}(y)}\notag\\
=&\langle x,y\rangle \frac{2l_0+N-2}{N-2} \mathcal{C}^{\frac{N-2}{2}}_{l_0}(\langle x,y \rangle)\notag\\
=&\frac{2l_0+N-2}{N-2}\cdot \frac{(l_0+1)\mathcal{C}^{\frac{N-2}{2}}_{l_0+1}(\langle x,y\rangle)+(l_0+N-3)\mathcal{C}^{\frac{N-2}{2}}_{l_0-1}(\langle x,y\rangle)}{2l_0+N-2}\notag\\
=&\frac{l_0+1}{N-2}\mathcal{C}^{\frac{N-2}{2}}_{l_0+1}(\langle x,y\rangle)+\frac{l_0+N-3}{N-2}\mathcal{C}^{\frac{N-2}{2}}_{l_0-1}(\langle x,y\rangle)\notag\\
=& \frac{l_0+1}{N-2}\cdot \frac{N-2}{2l_0+N}\sum_{m=1}^{d_{l_0+1}}Y_{l_0+1,m}(x)\overline{Y_{l_0+1,m}(y)}+\frac{l_0+N-3}{N-2}\cdot \frac{N-2}{2l_0+N-4}\sum_{m=1}^{d_{l_0-1}}Y_{l_0-1,m}(x)\overline{Y_{l_0-1,m}(y)},
\end{align}
where $\mathcal{C}$ is the Gegenbauer polynomial and we set $\mathcal{C}_{-1}^\lambda=0$. This computation is based on the addition theorem for spherical harmonics
$$
\sum_{m=1}^{d_l} Y_{l,m}(x)\overline{Y_{l,m}(y)}=\frac{2l+N-2}{N-2}\mathcal{C}^{\frac{N-2}{2}}_l,
$$
and the three-term recurrence relation of Gegenbauer polynomials
$$
(l+1)\mathcal{C}^\lambda_{l+1}(t)=2(l+\lambda) t \mathcal{C}_l^\lambda(t)-(l+2\lambda-1)\mathcal{C}_{l-1}^\lambda(t).
$$
See e.g. \cite{NIST, DLMF}. From (\ref{expansion}), we can recover the blocks corresponding to $\sum_{j=1}^N\mathcal{P}_{l_1}\mathcal{X}_j\mathcal{P}_{l_0}\mathcal{P}_{l_0}^*\mathcal{X}_j^*\mathcal{P}_{l_2}^*$. It's nonzero only for $l_1=l_2=l_0+1$ and $l_1=l_2=l_0-1$. Moreover, we have 
$$
\sum_{j=1}^N\mathcal{K}^j_{l_0,l_0+1}(x,y)=\frac{l_0+1}{N-2}\mathcal{C}^{\frac{N-2}{2}}_{l_0+1}(\langle x,y\rangle), \indent \sum_{j=1}^N\mathcal{K}^j_{l_0,l_0-1}(x,y)=\frac{l_0+N-3}{N-2}\mathcal{C}^{\frac{N-2}{2}}_{l_0-1}(\langle x,y\rangle),
$$
$$
\sum_{j=1}^N\mathcal{K}^j_{l_0+1,l_0-1}(x,y)=\sum_{j=1}^N\mathcal{K}^j_{l_0-1,l_0+1}(x,y)=0.
$$
Thus we can go back to (\ref{,ldk}) and we will have that
\begin{equation}
\label{BA}
\begin{aligned}
Z_{Villain}^N\sum_{j=1}^N M_{f_j}\leq &\frac{1}{L^{d-1}}\int\sum_{z_1'}\sum_{l_{z_1'}}b_{l_{z_1'}}\Bigg(\Big[-1+\Big(\frac{b_{l_{z_1'}-1}}{b_{l_{z_1'}}}\Big)^2\Big]\frac{l_{z_1'}+N-3}{N-2}\mathcal{C}^{\frac{N-2}{2}}_{l_{z_1'}-1}\Big(\langle \sigma_{(1,z_1')}, \sigma_{(2,z_1')}\rangle\Big)\\
&+\Big[-1+\Big(\frac{b_{l_{z_1'}+1}}{b_{l_{z_1'}}}\Big)^2\Big]\frac{l_{z_1'}+1}{N-2}\mathcal{C}^{\frac{N-2}{2}}_{l_{z_1'}+1}\Big(\langle \sigma_{(1,z_1')}, \sigma_{(2,z_1')}\rangle\Big)\Bigg)\prod_{\substack{z\sim w\\\{z,w\}\neq\{(1,z_1'),(2,z_1')\}}}h_\beta^{\mathbb{S}^{N-1}}(\sigma_z,\sigma_w)d\sigma\\
=&\frac{1}{L^{d-1}}\int\sum_{z_1'}\sum_{l_{z_1'}}\Bigg(\Big[-b_{l_{z_1'}+1}+\frac{b_{l_{z_1'}}^2}{b_{l_{z_1'}+1}}\Big]\frac{l_{z_1'}+N-2}{N-2}\mathcal{C}^{\frac{N-2}{2}}_{l_{z_1'}}\Big(\langle \sigma_{(1,z_1')}, \sigma_{(2,z_1')}\rangle\Big)\\
&+\Big[-b_{l_{z_1'}-1}+\frac{b_{l_{z_1'}}^2}{b_{l_{z_1'}-1}}\Big]\frac{l_{z_1'}}{N-2}\mathcal{C}^{\frac{N-2}{2}}_{l_{z_1'}}\Big(\langle \sigma_{(1,z_1')}, \sigma_{(2,z_1')}\rangle\Big)\Bigg)\prod_{\substack{z\sim w\\\{z,w\}\neq\{(1,z_1'),(2,z_1')\}}}h_\beta^{\mathbb{S}^{N-1}}(\sigma_z,\sigma_w)d\sigma.\\
\end{aligned}
\end{equation}
On the other hand, we have
\begin{equation}
\label{Ma}
\begin{aligned}
&Z_{Villain}^N\mathbb{E}_{Villain}^N \sum_{z_1'}\langle \sigma_{(1,z_1')}, \sigma_{(2,z_1')}\rangle\\
=&\int\sum_{z_1'}\langle \sigma_{(1,z_1')}, \sigma_{(2,z_1')}\rangle\prod_{z\sim w}h_\beta^{\mathbb{S}^{N-1}}(\sigma_z,\sigma_w)d\sigma\\
=&\int\sum_{z_1'}\sum_{l_{z_1'}}b_{l_{z_1'}}\Bigg(\frac{l_{z_1'}+N-3}{N-2}\mathcal{C}^{\frac{N-2}{2}}_{l_{z_1'}-1}\Big(\langle \sigma_{(1,z_1')}, \sigma_{(2,z_1')}\rangle\Big)+\frac{l_{z_1'}+1}{N-2}\mathcal{C}^{\frac{N-2}{2}}_{l_{z_1'}+1}\Big(\langle \sigma_{(1,z_1')}, \sigma_{(2,z_1')}\rangle\Big)\Bigg)\\
&\prod_{\substack{z\sim w\\\{z,w\}\neq\{(1,z_1'),(2,z_1')\}}}h_\beta^{\mathbb{S}^{N-1}}(\sigma_z,\sigma_w)d\sigma\\
=&\int\sum_{z_1'}\sum_{l_{z_1'}}\Bigg(b_{l_{z_1'}+1}\frac{l_{z_1'}+N-2}{N-2}\mathcal{C}^{\frac{N-2}{2}}_{l_{z_1'}}\Big(\langle \sigma_{(1,z_1')}, \sigma_{(2,z_1')}\rangle\Big)+b_{l_{z_1'}-1}\frac{l_{z_1'}}{N-2}\mathcal{C}^{\frac{N-2}{2}}_{l_{z_1'}}\Big(\langle \sigma_{(1,z_1')}, \sigma_{(2,z_1')}\rangle\Big)\Bigg)\cdot\\
&\prod_{\substack{z\sim w\\\{z,w\}\neq\{(1,z_1'),(2,z_1')\}}}h_\beta^{\mathbb{S}^{N-1}}(\sigma_z,\sigma_w)d\sigma.\\
\end{aligned}
\end{equation}
Inspired by the proof of the Villain model we want to compare the two equations above, note that $\frac{b_{l_{z_1'}}^2}{b_{l_{z_1'}+1}b_{l_{z_1'}-1}}=e^{\frac{1}{\beta}}$, we basically need to compare $(l_{z_1'}+N-2)b_{l_{z_1'}+1}+l_{z_1'}b_{l_{z_1'}-1}$ and  $e^{\frac{1}{\beta}}\Big((l_{z_1'}+N-2)b_{l_{z_1'}-1}+l_{z_1'}b_{l_{z_1'}+1}\Big)$. Note that
$$
\frac{b_{l_{z_1'}-1}}{b_{l_{z_1'}+1}}=e^{\frac{1}{2\beta}\Big[(l_{z_1'}+1)(l_{z_1'}+N-1)-(l_{z_1'}-1)(l_{z_1'}+N-3)\Big]}=e^{\frac{1}{\beta}(2l_{z_1'}+ N-2)},
$$
and we want to prove the following inequality
\begin{equation}
\label{sads}
\frac{l_{z_1'}+N-2}{2l_{z_1'}+N-2}e^{\frac{1}{\beta}(2l_{z_1'}+ N-2)}+\frac{l_{z_1'}}{2l_{z_1'}+N-2}e^0\leq e^{\frac{N-2}{\beta}}\Big(\frac{l_{z_1'}}{2l_{z_1'}+N-2}e^{\frac{1}{\beta}(2l_{z_1'}+ N-2)}+\frac{l_{z_1'}+N-2}{2l_{z_1'}+N-2}e^0\Big).    
\end{equation}
This looks like a Jensen-type inequality; to show this, we need to compare $\frac{1}{l_{z_1'}+N-2}(e^{\frac{1}{\beta}(2l_{z_1'}+2N-4)}-e^0)$ and $\frac{1}{l_{z_1'} }(e^{\frac{1}{\beta}(2l_{z_1'}+N-2)}-e^{\frac{1}{\beta}(N-2)})$ for $l_{z_1'}>0$ and it follows from the fact that $\frac{1}{u}(e^u-e^{-u})$ increases for $u>0$. From (\ref{sads}), we can see that
\[
 e^{\frac{1}{\beta}}\Big((l_{z_1'}+N-2)b_{l_{z_1'}-1}+l_{z_1'}b_{l_{z_1'}+1}\Big)\leq e^{\frac{N-1}{\beta}}\Big[(l_{z_1'}+N-2)b_{l_{z_1'}+1}+l_{z_1'}b_{l_{z_1'}-1}\Big].
\]
And going back to (\ref{Ma}) and (\ref{BA}), we will have that
$$
\sum_{j=1}^N M_{f_{j}}\leq \frac{1}{L^{d-1}}(e^{\frac{N-1}{\beta}}-1)\mathbb{E}_{Villain}^N\sum_{z_1'}\langle \sigma_{(1,z_1')},\sigma_{(2,z_1')} \rangle,
$$
if we rewrite the Gegenbauer polynomials in (\ref{BA}) and (\ref{Ma}) back into $\sum\limits_{m_{{z_1'}}=1}^{d_{l_{z_1'}}}Y_{l_{z_1'},m_{z_1'}}(x)\overline{Y_{l_{z_1'},m_{z_1'}}(y)}$, so that they become two diagonal infinite matrices and we can thus use the technique of reflection positivity in (\ref{OE}). So we finally get
$$
\sum_{j=1}^N M_{f_{j}}\leq e^{\frac{N-1}{\beta}}-1,
$$
which is not surprising, as we have a similar equation in the Villain model.  Combining with Lemma \ref{PDL}, we have 
$$
2-2\frac{1}{|\mathbb{T}_L^d|^2}\sum_{z,w}\mathbb{E}_{Villain}^N\langle\sigma_z,\sigma_w\rangle\leq\sum_{j=1}^N c\cdot M_{f_j}\leq c\cdot (e^{\frac{N-1}{\beta}}-1).
$$
And we can pick suitable $C^N_{Villain}<1$ and $\beta^N_M(d,C_{Villain})$ such that
$$
\frac{1}{|\mathbb{T}_L^d|^2}\sum_{z,w}\mathbb{E}_{Villain}^N\langle\sigma_z,\sigma_w\rangle \geq C^N_{Villain}
$$
for any $\beta>\beta^N_M$.
\end{proof}
\medskip
To check that the proof also works for $d\sigma_z$ replaced by any other finite positive measure on the sphere, we simply need to check the absolute integrability in (\ref{CSF}), as we did not use the condition that $d\sigma_z$ is uniform in the other part of the proof. And it still holds since
\begin{equation}
\begin{aligned}
&\sum_{l,m}\Big(\prod_k p(l_k)\Big)\cdot \Big(\int_{X^{{L}^{d-1}}} \prod_k[|Y_{l_k,m_k}(x_k)|^2d\mu(x_k)]\Big)\\
=&\sum_{l}\Big(\prod_k p(l_k)\Big)\cdot \Big(\int_{X^{{L}^{d-1}}}\sum_{m} \prod_k[|Y_{l_k,m_k}(x_k)|^2d\mu(x_k)]\Big)\\
=&\sum_{l}\Big(\prod_k p(l_k)\Big)\cdot \Big(\int_{X^{{L}^{d-1}}} \prod_k[\sum_{m_k=1}^{d_{l_k}}|Y_{l_k,m_k}(x_k)|^2d\mu(x_k)]\Big)\\
=&\sum_l\Big(\prod_k p(l_k) d_{l_k}\Big)\cdot\mu(X)^{L^{d-1}},
\end{aligned}
\end{equation}
which is finite. In the last equality, we used the fact that (see \cite{DaiXu}, Corollary 1.2.7)
$$
\sum_{m_k=1}^{d_{l_k}}|Y_{l_k,m_k}(x_k)|^2=d_{l_k}.
$$
This equation is also known as the Unsöld theorem for $N=3$ in the physics community.

Now we focus on general continuous rotation-invariant kernels $K(x,y)=K_0(x\cdot y)\in L^2(\mathbb{S}^{N-1}\otimes\mathbb{S}^{N-1}, \mathbb{R})$ on the sphere with $K>0$. Recall (\ref{RI}) that it can be written in the form
\begin{equation}
K(x,y)=\sum_{l=0}^\infty q_l\sum_{m=1}^{d_l}Y_{l,m}(x)\overline{Y_{l,m}(y)}.
\end{equation}
To show that this proof is suitable for any positive sequence $\{q_l\}$ satisfying
\[
\sup_{l\geq1}\Big|\log\frac{q_l^2}{q_{l-1}q_{l+1}}\Big|\to 0 \text{ and } \Big|\log\frac{q_1}{q_0}\Big|\to 0,
\] 
as $\beta \to \infty$, we notice that in the proof we only need the value of $b_l$ when we check the absolute integrability and we estimate (\ref{Ma}). For absolute integrability, we note that
\begin{equation}
K(x,x)=\sum_{l=0}^\infty q_l\sum_{m=1}^{d_l}Y_{l,m}(x)\overline{Y_{l,m}(x)}=\sum_{l=0}^{\infty}q_l d_l<\infty,
\end{equation}
which is one of the inequalities we need. For the absolute integrability of $q_l$ replaced by $\frac{q_{l_v\pm 1}}{q_{l_v}}\cdot q_{l_v}$ or $\Big(\frac{q_{l_v\pm 1}}{q_{l_v}}\Big)\Big(\frac{q_{l_v\pm 1}}{q_{l_v}}\Big)\cdot q_{l_v}$, note that $\frac{q_{l_v\pm 1}}{q_{l_v}}$, $\frac{q_{l_v\pm 1+1}}{q_{l_v+1}}$ and $\frac{q_{l_v\pm 1-1}}{q_{l_v-1}}$ are very close to each other since
\[
\sup_{l\geq1}\Big|\log\frac{q_l^2}{q_{l-1}q_{l+1}}\Big|\to 0.
\]
So we only need to consider the case where $q_l$ is replaced by $q_{l-2}$,$q_{l-1}$, $q_{l+1}$ and $q_{l+2}$. For this, note that
\[
\sum_{l=0}^{\infty}q_{l} d_{l-2}\leq\sum_{l=0}^{\infty}q_{l} d_{l-1}\leq\sum_{l=0}^{\infty}q_{l} d_l<\infty
\]
and
\[
\frac{1}{N^2}\sum_{l=0}^{\infty}q_{l} d_{l+2}\leq \frac{1}{N}\sum_{l=0}^{\infty}q_{l}d_{l+1}\leq\sum_{l=0}^{\infty}q_{l} d_l<\infty
\]
since $d_l\leq d_{l+1}\leq N d_{l}$ and $q_l>0$.

About the proof of (\ref{Ma}) we only need the fact that $\log\frac{b_{l_{z_1'}}^2}{b_{l_{z_1'}+1}b_{l_{z_1'}-1}}\to 0$ and $\frac{1}{2l_{z_1'}+N-2}\log\frac{b_{l_{z_1'}-1}}{b_{l_{z_1'}+1}}\to 0$ uniformly when $\beta\to\infty$, and they also hold in this condition.

We wish to point out that $a_{l,m}^\pm$ and $Y_{l,m}$ are exactly $r_{mn}$ and $\phi_n$ in Lemma \ref{PDL1}. Thus, the computation in this theorem is a more detailed analysis of the spectrum of the Hilbert-Schmidt operators.
\appendix 

\section{Spherical harmonics}
\label{APA}
\begin{lem}
\label{SPS}
For $N\geq 3$, there exist spherical harmonics $\{Y_{l,m}\}_{l\geq0,1\leq m\leq d_l}(d_l=\frac{(2l+N-2)(l+N-3)!}{l! (N-2)!})$, such that for any $l\geq0$, $\{Y_{l,m}\}_{1\leq m\leq d_l}$ is an orthonormal basis of spherical harmonics of degree $l$ on $\mathbb{S}^{N-1}$, and for any $x=(x^1,x^2,\dots,x^N)\in \mathbb{S}^{N-1}$ we have
\begin{equation}
x^N \cdot Y_{l,m}(x)=a^{-}_{l,m}Y_{l-1,m}(x)+a^{+}_{l,m}Y_{l+1,m}(x),
\end{equation}
$a_{l,m}^{-}, a_{l,m}^+$ are real constants that depend only on $l,m$ and we set $Y_{l,m}=0$ if $m>d_l$ or $l<0$.
\end{lem}
\begin{proof}
Let $x^N=\cos\theta$ and $(x^{1},\dots,x^{N-1})= \omega\cdot\sin\theta$, where $\omega \in \mathbb{S}^{N-2}$. A typical way to define $\{Y_{l,m}\}$ recursively is as follows. If $\{\Tilde{Y}_{k,n}\}$ is an orthonormal basis for spherical harmonics of degree $k$ on $\mathbb{S}^{N-2}$ for any $k\geq 0$, we consider the following functions:
\[
Y_{l,k,n}(x):=c^A_{l,k} \sin^k \theta \cdot\mathcal{C}_{l-k}^{k+\frac{N-2}{2}}(\cos\theta)\Tilde{Y}_{k,n}(\omega),
\]
where $c^A>0$ are the normalization constants, $\mathcal{C}$ are Gegenbauer polynomials, and $k\leq l$. Then $\{Y_{l, m(k,n)}=Y_{l,k,n}\}$ form the orthonormal basis of spherical harmonics of degree $l$ on $\mathbb{S}^{N-1}$, where we relabel the functions by a bijection $m$ onto $\{1,\dots,d_l\}$. Note that we can find a suitable way to relabel them so that $m(k,n)$ does not depend on $l$. For $N-2=1$, we can take $\Tilde{Y}_{k,n}(\omega)\propto(\mathds{1}_{n=1}\cos k\omega+\mathds{1}_{n=2}\sin k\omega)$ as the orthonormal basis for $\mathbb{S}^1$, and this coincides with the definition in Theorem 1.5.1 of \cite{DaiXu}. Now, by the three-term recurrence of Gegenbauer polynomials 
\begin{equation}
t\,\mathcal{C}^{u}_{j}(t)=\frac{j+1}{2(j+u)}\,\mathcal{C}^{u}_{j+1}(t)+\frac{j+2u-1}{2(j+u)}\,\mathcal{C}^{u}_{j-1}(t),
\end{equation}
where $j\geq 0$, $u\geq 0$ and $\mathcal{C}_{-1}^u:=0$
we have
\begin{equation}
\cos\theta \,\mathcal{C}^{u}_{j}(\cos\theta)=\frac{j+1}{2(j+u)}\,\mathcal{C}^{u}_{j+1}(\cos\theta)+\frac{j+2u-1}{2(j+u)}\,\mathcal{C}^{u}_{j-1}(\cos\theta).
\end{equation}
Let $j=l-k$ and $u=k+\frac{N-2}{2}$, we get 
\begin{equation}
x^N \cdot Y_{l,k,n}(x)=c_{l,k}^B\,Y_{l+1,k,n}(x)+c_{l,k}^C\,Y_{l-1,k,n}(x),
\end{equation}
where we take $c_{l,k}^A$ into the equations to get the constants $c_{l,k}^B$, $c_{l,k}^C$. By the bijection $m(k,n)$ we have real constants $a_{l,m}^+$, $a_{l,m}^-$ such that
\begin{equation}
x^N \cdot Y_{l,m}(x)=a^{-}_{l,m}Y_{l-1,m}(x)+a^{+}_{l,m}Y_{l+1,m}(x).
\end{equation}
\end{proof}
\section{Some alternative proofs}
\label{APB}
For convenience, we have followed the framework of the appendix of \cite{FSS}. In this section, we provide two alternative proofs for the long-range order of the spin $O(N)$ model under another framework, which also appeared in Section 3 of \cite{Ondual}. The first proof uses Gaussian domination, and the second uses the transfer matrix formalism in Lemma \ref{PDL}, but the functions are very close. From our perspective, these two approaches, together with the approach in \cite{Ondual}, are somehow equivalent.

We first recall the setting of the spin $O(N)$ model as a special case of Theorem \ref{INF0}, where $\sigma: \mathbb{T}_L^d \to \mathbb{R}^N $($d \geq 3$) is a random configuration with Hamiltonian
\[
H(\sigma) = \sum_{z\sim w,\, j=1 \dots N} V(\sigma_z^j - \sigma_w^j)
\]
and probability measure
\[
\mathbb{P}_V(d\sigma)\propto \exp\big(-H(\sigma)\big)\prod_zd\nu(\sigma_z).
\]
For the spin $O(N)$ model, $V(x)=\frac{\beta}{2}x^2$, $\nu$ is the normalized surface measure on the sphere.
\begin{thm}
\label{B.1}
We have long-range order of the spin $O(N)$ model on torus of dimension $d\geq 3$ in the low-temperature regime: there exist constants $C_{Spin}>0$ and $\beta_{Spin}(C_{Spin},N,d)>0$ such that for any $\beta>\beta_{Spin}$ and $z,w\in \mathbb{T}_L^d$ we have
\begin{equation}
\mathbb{E}_{V}\langle \sigma_z, \sigma_w\rangle>C_{Spin}.
\end{equation}
\end{thm}
\begin{proof}
We take $p(t):=\frac{1}{\sqrt{2\pi \beta}}\exp(-\frac{t^2}{2\beta})$ so that $\int p(t)\, e^{itx}\, dt = e^{-V(x)}$. We do the trick of (\ref{int1}) on the whole lattice. Note that 
for any skew-symmetric function $R:\mathbb{T}_L^d\times\mathbb{T}_L^d\to \mathbb{C}$,
\begin{align}
\label{B.2}
&\Big|\sum_{z \sim w} R(z,w)(\sigma^N_z -\sigma^N_w)\Big|^2 \exp\big(-\sum_{\substack{z\sim w\\j=1,\dots,N}}V(\sigma^j_z-\sigma^j_w)\big)\notag\\
=&\Big|\sum_{z \sim w} R(z,w)(\sigma^N_z -\sigma^N_w)\Big|^2 \prod_{\substack{z\sim w\\j=1,\dots,N}} \int_{t^j_{z,w}}p(t^j_{z,w})e^{it^j_{z,w}(\sigma_z^j-\sigma_w^j)}dt^j_{z,w}\notag\\
=&\int_t \Big|\sum_{z \sim w} R(z,w)(\sigma^N_z -\sigma^N_w)\Big|^2 \prod_{\substack{z\sim w\\j=1,\dots,N}} \Big( p(t^j_{z,w})e^{it^j_{z,w}(\sigma_z^j-\sigma_w^j)}\Big) dt\notag \\
=&-\int_t\Bigg[\Big(\sum_{z \sim w} R(z,w)\partial_{t_{z,w}^N}\Big)\Big(\sum_{z \sim w} \overline{R(z,w)}\partial_{t_{z,w}^N}\Big)\Big(\prod_{\substack{z\sim w\\j=1,\dots,N}}p(t^j_{z,w})\Big)\Bigg]\Big(\prod_{\substack{z\sim w\\j=1,\dots,N}}e^{it^j_{z,w}(\sigma_z^j-\sigma_w^j)}\Big)dt\notag\\
=&-\int_t\Big(\sum_{\substack{z_1\sim w_1\\z_2\sim w_2}}\mathds{1}_{\{z_1,w_1\}\neq\{z_2,w_2\}}R(z_1,w_1)\overline{R(z_2,w_2)}\frac{p'(t_{z_1,w_1}^N)}{p(t_{z_1,w_1}^N)}\frac{p'(t_{z_2,w_2}^N)}{p(t_{z_2,w_2}^N)}+\sum_{z\sim w}|R(z,w)|^2\frac{p''(t_{z,w}^N)}{p(t_{z,w}^N)}\Big)\notag\\
&\Big(\prod_{\substack{z\sim w\\j=1,\dots,N}}p(t^j_{z,w})\Big)\Big(\prod_{\substack{z\sim w\\j=1,\dots,N}}e^{it^j_{z,w}(\sigma_z^j-\sigma_w^j)}\Big)dt\\
=&-\int_t\Big[\Big|\sum_{z\sim w}R(z,w)\frac{p'(t_{z,w}^N)}{p(t_{z,w}^N)}\Big|^2+\sum_{z\sim w}|R(z,w)|^2\frac{p''(t_{z,w}^N)}{p(t_{z,w}^N)}-\sum_{z\sim w}|R(z,w)|^2\Big(\frac{p'(t_{z,w}^N)}{p(t_{z,w}^N)}\Big)^2\Big]\Big(\prod_{\substack{z\sim w\\j=1,\dots,N}}p(t^j_{z,w})e^{it^j_{z,w}(\sigma_z^j-\sigma_w^j)}\Big)dt.\notag
\end{align}
For the Gaussian law, it is not hard to check the absolute integrability, so that we can interchange the order of summation. Note that $\frac{p''}{p}-(\frac{p'}{p})^2=(\log p)''=-\frac{1}{\beta}$, thus 
\begin{equation}
\begin{aligned}
&-\int_t\Big[\sum_{z\sim w}|R(z,w)|^2\frac{p''(t_{z,w}^N)}{p(t_{z,w}^N)}-\sum_{z\sim w}|R(z,w)|^2\Big(\frac{p'(t_{z,w}^N)}{p(t_{z,w}^N)}\Big)^2\Big]\Big(\prod_{\substack{z\sim w\\j=1,\dots,N}}p(t^j_{z,w})e^{it^j_{z,w}(\sigma_z^j-\sigma_w^j)}\Big)dt\\
=&\frac{1}{\beta}\int_t\Big[\sum_{z\sim w}|R(z,w)|^2\Big]\Big(\prod_{\substack{z\sim w\\j=1,\dots,N}}p(t^j_{z,w})e^{it^j_{z,w}(\sigma_z^j-\sigma_w^j)}\Big)dt.
\end{aligned}
\end{equation}
Now for any $z_0, w_0 \in \mathbb{T}_L^d$, we want to pick some $R$ such that $\sum\limits_{z \sim w} R(z,w)(\sigma^N_z -\sigma^N_w)=\sigma^N_{z_0}-\sigma^N_{w_0}$ so that we have computed $(\sigma^N_{z_0}-\sigma^N_{w_0})^2 \exp\big(-\sum\limits_{\substack{z\sim w\\ j=1,\dots, N}}V(\sigma^j_z-\sigma^j_w)\big)$. A natural choice is $R(z,w):=h(z)-h(w)$, where $h:\mathbb{T}_L^d\to \mathbb{R}$ satisfies $\Delta h=\delta_{z_0}-\delta_{w_0}$. In this case $\sum\limits_{z\sim w} |R(z,w)|^2=\sum\limits_{z\sim w}(h(z)-h(w))^2=\sum\limits_z h(z)\Delta h(z)=h(z_0)-h(w_0)$. This is upper bounded by some constant $C_B(d)$ related to Green's function on the torus. For example, an explicit $h$ is given in \cite{Mywork}. So, we can pick suitable $C_{Spin}<1$ and $\beta_{Spin}(C_{Spin},N,d)$ such that for any $\beta>\beta_{Spin}$ and $z_0,w_0\in \mathbb{T}_L^d$ we have
\begin{equation}
\begin{aligned}
&-\int_t\sum_{j=1}^N\Big[\sum_{z\sim w}|R(z,w)|^2\frac{p''(t_{z,w}^j)}{p(t_{z,w}^j)}-\sum_{z\sim w}|R(z,w)|^2\Big(\frac{p'(t_{z,w}^j)}{p(t_{z,w}^j)}\Big)^2\Big]\Big(\prod_{\substack{z\sim w\\j=1,\dots,N}}p(t^j_{z,w})e^{it^j_{z,w}(\sigma_z^j-\sigma_w^j)}\Big)dt\\
\leq&\frac{N}{\beta}C_B(d)\int_t\Big(\prod_{\substack{z\sim w\\j=1,\dots,N}}p(t^j_{z,w})e^{it^j_{z,w}(\sigma_z^j-\sigma_w^j)}\Big)dt\\
=&\frac{N}{\beta}C_B(d)\exp\big(-\sum_{\substack{z\sim w\\ j=1,\dots, N}}V(\sigma^j_z-\sigma^j_w)\big)\\
\leq&(2-2C_{Spin})\exp\big(-\sum_{\substack{z\sim w\\ j=1,\dots, N}}V(\sigma^j_z-\sigma^j_w)\big).
\end{aligned}
\end{equation}
By (\ref{B.2}), to show that $\mathbb{E}_{V} \langle\sigma_{z_0},\sigma_{w_0}\rangle>C_{Spin}$ we only need to prove that 
\begin{equation}
\label{B.6}
\int_\sigma\int_t\Big|\sum_{z\sim w}R(z,w)\frac{p'(t_{z,w}^N)}{p(t_{z,w}^N)}\Big|^2\Big(\prod_{\substack{z\sim w\\j=1,\dots,N}}p(t^j_{z,w})e^{it^j_{z,w}(\sigma_z^j-\sigma_w^j)}\Big)dtd\sigma\geq 0.
\end{equation}
 
Note that $R(z,w)=h(z)-h(w)$, $\frac{p'(t_{z,w}^N)}{p(t_{z,w}^N)}=-\frac{t_{z,w}^N}{\beta}$, this is equivalent to 
\begin{equation}
\label{..}
\int_\sigma\int_t\Big(\sum_{z\sim w}\Big(h(z)-h(w)\Big)t_{z,w}^N\Big)^2\Big(\prod_{\substack{z\sim w\\j=1,\dots,N}}p(t^j_{z,w})e^{it^j_{z,w}(\sigma_z^j-\sigma_w^j)}\Big)dt\geq 0, 
\end{equation}
which will be nothing but
\[
-\partial^2_\eta\int_\sigma\int_t\Big(\prod_{\substack{z\sim w\\j=1,\dots,N}}p(t^j_{z,w})e^{it^j_{z,w}\big(\sigma_z^j-\sigma_w^j+\mathds{1}_{j=N}(h(z)-h(w))\eta\big)}\Big)dtd\sigma\Big|_{\eta=0}\geq 0.
\]
This is 
\[
\partial^2_\eta\int_\sigma\exp\big(-\sum_{\substack{z\sim w\\ j=1,\dots, N}}V(\sigma^j_z-\sigma^j_w+\mathds{1}_{j=N}(h(z)-h(w))\eta)\big)d\sigma\Big|_{\eta=0}\leq 0.
\]
Note that if we take $H_0(z):=(0,0,\dots,\eta h(z))$, this follows from the so-called Gaussian domination \cite{FSS}:
\[
\int_\sigma\exp\big(-\sum_{\substack{z\sim w\\ j=1,\dots,  N}}V(\sigma^j_z-\sigma^j_w+H_0(z)^j-H_0(w)^j)\big)d\sigma\leq \int_\sigma\exp\big(-\sum_{\substack{z\sim w\\ j=1,\dots, N}}V(\sigma^j_z-\sigma^j_w)\big)d\sigma,
\]
and the symmetry under $\eta \to -\eta$.
\end{proof}
\begin{thm}
\label{B}
We have long-range order of the spin $O(N)$ model on torus of dimension $d\geq 3$ in the low-temperature regime: there exist constants $C_{Spin}>0$ and $\beta_{Spin}(C_{Spin},N,d)>0$ such that for any $\beta>\beta_{Spin}$ and $z,w\in \mathbb{T}_L^d$ we have
\begin{equation}
\frac{1}{|\mathbb{T}_L^d|^2}\sum\limits_{z, w}\mathbb{E}_V\langle\sigma_z, \sigma_w\rangle >C_{Spin}.
\end{equation}
\end{thm}
\begin{proof}
For any $\omega=(\omega_1,\dots,\omega_d)\in\{1,\dots,L\}^d$, we take $R(z,w)=\mathds{1}_{\{z-w=e_d\}}(h(z)-h(w))$ in (\ref{B.2}), where $e_j\in \mathbb{R}^d$ denotes the $j$-th standard basis vector (with \(1\) in the \(j\)-th entry and \(0\) elsewhere), and $h(z)=e^{2\pi i\cdot \frac{\omega \cdot z}{L}}$. In this case
$$
\Big|\sum_{z\sim w}R(z,w)(\sigma_z^N-\sigma_w^N)\Big|^2=\Big|\sum_{z}\big(2h(z)-h(z-e_d)-h(z+e_d)\big)\sigma_z^N\Big|^2=\Big|2\Big(1-\cos(2\pi\cdot \frac{\omega_d}{L})\Big)\sum_{z}h(z)\sigma_z^N\Big|^2.
$$
On the other hand,
$$
\sum_{z\sim w} |R(z,w)|^2=\sum_z\overline{\Big(2h(z)-h(z-e_d)-h(z+e_d)\Big)}h(z)=2\Big(1-\cos(2\pi\cdot \frac{\omega_d}{L})\Big)\sum_{z}|h(z)|^2=2\Big(1-\cos(2\pi\cdot \frac{\omega_d}{L})\Big)L^d.
$$
So, by (\ref{B.2}), if we have
\begin{equation}
 \label{jjjj}
\int_\sigma\int_t\Big|\sum_{z\sim w}R(z,w)\frac{p'(t_{z,w}^N)}{p(t_{z,w}^N)}\Big|^2\Big(\prod_{\substack{z\sim w\\j=1,\dots,N}}p(t^j_{z,w})e^{it^j_{z,w}(\sigma_z^j-\sigma_w^j)}\Big)dtd\sigma\geq 0,
\end{equation}
we already get the infrared bound
$$
\mathbb{E}_{V}\Big|2\Big(1-\cos(2\pi\cdot \frac{\omega_d}{L})\Big)\sum_{z}h(z)\sigma_z^N\Big|^2\leq M \cdot2\Big(1-\cos(2\pi\cdot \frac{\omega_d}{L})\Big)L^d,
$$
which is just (\ref{lalala}) if we take $f(\sigma_z)=\sigma_z^N$ there. Note that we omit the reason that we can replace $-\frac{p''}{p}+(\frac{p'}{p})^2$ with $M$ and the check of the absolute integrability, which is similar to the part in Theorem \ref{INF}. And by the same argument in Lemma \ref{PDL} we have
$$
\frac{1}{|\mathbb{T}_L^d|^2}\sum\limits_{z, w}\mathbb{E}_V|\sigma_z - \sigma_w|^2 = \frac{N}{|\mathbb{T}_L^d|^2}\sum\limits_{z, w}\mathbb{E}_V|\sigma_z^N - \sigma_w^N|^2\leq c(d)NM=    C(N, d)M ,
$$
if we take $C(N,d)=Nc(d)$.
It remains to prove (\ref{jjjj}), 
and we consider $\sigma_k:\mathbb{T}_L^{d-1}\to \mathbb{R}^N$ such that $\sigma_k(\cdot)=\sigma_{(\cdot,k)}$ with measure $$d\sigma_k:=\prod\limits_{\substack{z',w'\in \mathbb{T}_L^{d-1}\\z'\sim w'}} e^{-\sum\limits_{j=1}^NV(\sigma_k(z')^j-\sigma_k(w')^j)}\prod\limits_{z'\in\mathbb{T}_L^{d-1}}d\sigma_{(z',k)},$$
and $t_k:\mathbb{T}_L^{d-1}\to \mathbb{R}^N$ such that $t_k(\cdot)=t_{(\cdot,k),(\cdot,k+1)}$ with measure
$$
dt_k:=\prod_{z'\in \mathbb{T}_L^{d-1}}\prod_{j=1}^N p(t_{(z',k),(z',k+1)}^j) dt_{(z',k),(z',k+1)}^j.
$$ We can rewrite (\ref{jjjj}) since 
\begin{equation}
\begin{aligned}
&\int_\sigma\int_t\Big|\sum_{z\sim w}R(z,w)\frac{p'(t_{z,w}^N)}{p(t_{z,w}^N)}\Big|^2\Big(\prod_{\substack{z\sim w\\j=1,\dots,N}}p(t^j_{z,w})e^{it^j_{z,w}(\sigma_z^j-\sigma_w^j)}\Big)dtd\sigma\\
=&\int_\sigma\int_{t}\Big|\sum_{z-w=e_d}R(z,w)\frac{p'(t_{z,w}^N)}{p(t_{z,w}^N)}\Big|^2\Big(\prod_{\substack{z- w=e_d\\j=1,\dots,N}}p(t^j_{z,w})e^{it^j_{z,w}(\sigma_z^j-\sigma_w^j)}\Big)\cdot \Big(\prod_{\substack{z\sim w\\z- w\neq e_d\\j=1,\dots,N}}p(t^j_{z,w})e^{it^j_{z,w}(\sigma_z^j-\sigma_w^j)}\Big)dtd\sigma\\
=&\int_{\sigma_1}\dots\int_{\sigma_{L}}\int_{t_1}\dots\int_{t_L}\Big|\sum_{k=1}^L\sum_{z'\in\mathbb{T}_L^{d-1}}R((z',k),(z',k+1))\frac{p'(t_k(z')^N)}{p(t_k(z')^N)}\Big|^2
\prod_{k=1}^L\Big(e^{it_k\cdot(\sigma_k-\sigma_{k+1})}dt_kd\sigma_k\Big)\\
=&\int_{\sigma_1}\dots\int_{\sigma_{L}}\int_{t_1}\dots\int_{t_L}\Big|\sum_{k=1}^L\sum_{z'\in\mathbb{T}_L^{d-1}}R((z',k),(z',k+1))\frac{p'(t_k(z')^N)}{p(t_k(z')^N)}\Big|^2
\prod_{k=1}^L\Big(e^{i\sigma_k\cdot(t_k-t_{k-1})}dt_kd\sigma_k\Big).\\
\end{aligned}
\end{equation}
Now if we take $f_0(t_k):=\sum\limits_{z'\in\mathbb{T}_L^{d-1}}R((z',1),(z',2))\frac{p'(t_k(z')^N)}{p(t_k(z')^N)}$, and $K_1(t_k,t_{k-1}):=\int_{\sigma_k}e^{i\sigma_k(t_k-t_{k-1})}d\sigma_k$, we can furthermore rewrite (\ref{jjjj}) into
$$
\int_{t_1}\dots\int_{t_L}\Big|\sum_{k=1}^L e^{2\pi i\cdot\frac{(k-1)\omega_d}{L}}f_0(t_k)\Big|^2K_1(t_1,t_2)\dots K_1(t_L,t_1)d_{t_1}\dots d_{t_L}\geq 0.
$$
Note that $d_{t_k}$ is a finite measure, $f_0$ is an $L^2$ function (we check it in Theorem \ref{INF}), and $K_1$ is bounded since $d\sigma_k$ is a finite measure; (\ref{PD2}) can also be checked since $K_1$ is positive definite; we can use all the techniques in the proof of Lemma \ref{PDL} and simply need to check the inequality
\begin{equation} 
 \sum_{j=1}^{L} (\lambda_n^{L-j+1}\lambda_m^{j-1}+\lambda_{m}^{L-j+1}\lambda_n^{j-1} ) e^{2\pi i\cdot \frac{(j-1)\omega_d}{L}} 
\geq 0,
\end{equation}
for $\lambda_n,\lambda_m\geq 0$. As computed in Lemma A.2 of \cite{FSS},  for $\lambda_m\geq\lambda_n$ and $\lambda_m>0$, the left side is 
\[
\lambda_m^L(1 - (\frac{\lambda_n}{\lambda_m})^L)\!(1 -  (\frac{\lambda_n}{\lambda_m} )^2 )
\cdot  |1 - e^{2\pi i\frac{\omega_d}{L}} \cdot \frac{\lambda_n}{\lambda_m} |^{-2},
\]
which is non-negative. If $\lambda_m=\lambda_n=0$ or $\lambda_n>\lambda_m$, it is also clear.
\end{proof}
\medskip

Theorem \ref{B.1} is slightly stronger and has also been proved in \cite{Mywork}. 

In this framework, we can see that the main difficulty in these problems is to prove (\ref{B.6}). Note that if we define the Fourier transform of the measure $d\sigma_z$ by $\mathscr{R}(x):=\int e^{i\sigma_z\cdot x }d\sigma_z$, then (\ref{B.6}) is
\begin{equation}
\int_t\Big|\sum_{z\sim w}R(z,w)\frac{p'(t_{z,w}^N)}{p(t_{z,w}^N)}\Big|^2\prod_{\substack{z\sim w\\j=1,\dots,N}}p(t^j_{z,w})\prod_z \mathscr{R}(\sum_{j=1}^dt_{z,z+e_j}-\sum_{j=1}^dt_{z-e_j,z})dt\geq 0.
\end{equation}
If $\mathscr{R}\geq 0$, this inequality is trivial. In this case, $\prod\limits_{\substack{z\sim w}}p(t^j_{z,w})\prod\limits_z \mathscr{R}(\sum\limits_{j=1}^dt_{z,z+e_i}-\sum\limits_{j=1}^dt_{z-e_i,z})$ can be seen as a density function, and we can get a model of $t$ on the edges. In Section 3 of \cite{Ondual}, this technique is applied when $d\sigma$ is the counting measure on $\mathbb{Z}$, and we ask $t$ to take values in $(-\pi, \pi]$. Under this Fourier duality, (\ref{B.6}) can even be proved on any finite graph. 

We also point out that the Gaussian domination approach seems only suitable for the Gaussian law, in which case $\frac{p'(x)}{p(x)}\propto x$, and in the dual space this will become the derivative. The author is not aware of whether we could prove the result for other $p$ if it is not related to the derivative. As a counterexample to Gaussian domination for the spin $O(N)$ model on a non-torus graph is given in \cite{Mywork}, both proofs in Theorem \ref{B.1} and Theorem \ref{B} seem to be restricted to the torus.

\end{document}